\documentclass{jfm}


\usepackage{graphicx}
\usepackage{newtxtext}
\usepackage{newtxmath}
\usepackage{ragged2e}
\usepackage{blindtext}
\usepackage{natbib}
\usepackage{xcolor}
\usepackage{hyperref}

\hypersetup{
colorlinks = true,
urlcolor   = blue,
citecolor  = blue,
}
\DeclareUnicodeCharacter{2009}{\,} 

\usepackage{caption}

\newcommand{\RomanNumeralCaps}[1]
\linenumbers
\usepackage[format=plain]{caption}

\makeatletter
\patchcmd{\NAT@citex}
  {\@citea\NAT@hyper@{%
     \NAT@nmfmt{\NAT@nm}%
     \hyper@natlinkbreak{\NAT@aysep\NAT@spacechar}{\@citeb\@extra@b@citeb}%
     \NAT@date}}
  {\@citea\NAT@nmfmt{\NAT@nm}%
   \NAT@aysep\NAT@spacechar\NAT@hyper@{\NAT@date}}{}{}

\patchcmd{\NAT@citex}
  {\@citea\NAT@hyper@{%
     \NAT@nmfmt{\NAT@nm}%
     \hyper@natlinkbreak{\NAT@spacechar\NAT@@open\if*#1*\else#1\NAT@spacechar\fi}%
       {\@citeb\@extra@b@citeb}%
     \NAT@date}}
  {\@citea\NAT@nmfmt{\NAT@nm}%
   \NAT@spacechar\NAT@@open\if*#1*\else#1\NAT@spacechar\fi\NAT@hyper@{\NAT@date}}
  {}{}

\makeatother

\title{Viscoelastic control of acoustic particle migration and trapping in microchannels} 

\author{T. Sujith\aff{}\aff{} \and  A. K. Sen\aff{}
 \corresp{\email{ashis@iitm.ac.in}}}

\affiliation{\aff{}Micro Nano Bio Fluidics Unit, Department of Mechanical Engineering, Indian Institute of Technology Madras, Chennai-600036, India 
}

\begin{document}
\maketitle

\begin{abstract}

Particle migration and trapping in ultrasonically actuated microscale flows arise from the competition between acoustic radiation forces and streaming-induced drag. While these mechanisms are well understood in Newtonian fluids, the role of fluid viscoelasticity in governing particle dynamics remains largely unexplored. Here, we investigate particle transport and trapping in a viscoelastic fluid within an ultrasonically excited microchannel under the combined action of acoustic streaming and radiation forces. Using a perturbation framework, we solve the continuity, momentum and constitutive equations for an Oldroyd-B fluid to obtain the oscillatory acoustic field and the resulting steady streaming flows in the bulk and near-wall boundary layers. Acoustic radiation forces, incorporated through a semi-analytical model, drives particle migration, while streaming-induced drag can oppose, alter or suppress trapping. We show that particle trajectories and equilibrium trapping locations are governed primarily by the Deborah number ($De$) and viscous diffusion number ($Dv$). At high $Dv$, increasing $De$ shifts the trapping location from the bulk region to the channel wall, pressure nodal line, channel centre or ultrasound symmetry line. We further determine the critical particle size governing the transition between radiation-dominated and streaming-dominated regimes as a function of $De$ and $Dv$. The critical particle size can become significantly smaller than that in a Newtonian fluid, enabling effective manipulation of submicron particles and overcoming a key limitation of conventional acoustofluidics. These results demonstrate how viscoelasticity fundamentally modifies acoustophoretic transport and establish new mechanisms for tunable particle migration and trapping in complex fluids.
\end{abstract}

\section{Introduction}
\label{sec:1}

Acoustofluidics has emerged as a powerful approach for manipulating suspended particles and cells in microfluidic systems, with applications ranging from particle focusing and separation to pumping and mixing \citep{Petersson2007, Sajeesh2014}. When an acoustic field interacts with suspended particles, momentum transfer from the sound wave generates acoustic radiation forces (ARF) that drive particles towards pressure nodes or antinodes depending on their acoustic contrast \citep{King1934,Gorkov1962}. Simultaneously, momentum transfer to the fluid produces steady acoustic streaming flows through gradients in the time-averaged momentum flux \citep{Rayleigh1884}, generating drag forces that can transport suspended particles and enhance mixing in confined geometries \citep{Wiklund2012}. The relative importance of these two mechanisms strongly depends on particle size: in microchannels driven at MHz frequencies, radiation forces typically dominate the motion of larger particles, whereas streaming-induced drag becomes increasingly important for smaller particles such as bacteria, viruses and exosomes \citep{Lenshof2012, Wiklund2012, Hahn2015-zu,Hammarstrm2012}. The competition between radiation forces and streaming therefore fundamentally governs particle migration and trapping dynamics in acoustofluidic systems. Here, we demonstrate that fluid viscoelasticity provides a new mechanism for actively modulating this competition and thereby controlling particle trapping.

Under acoustic excitation, suspended particles migrate towards equilibrium locations corresponding to extrema of the effective acoustic potential, where the net force vanishes \citep{Evander2012}. These trapping locations are determined by the balance between acoustic radiation forces and streaming-induced drag, making their interplay central to predicting and controlling particle trajectories in microchannels. The drag force originates from acoustic streaming, a steady second-order flow generated through viscous dissipation of acoustic energy when sound waves interact with fluid boundaries. First observed experimentally by \cite{Faraday1831,Kundt1874}, and later explained theoretically by \cite{Rayleigh1884}, acoustic streaming can arise either from dissipation in the bulk fluid (commonly referred to as quartz wind) or from viscous boundary layers near solid interfaces. In confined microfluidic systems, boundary-driven streaming is typically the dominant mechanism and therefore constitutes the focus of the present study.

Particle migration in acoustofluidics is highly sensitive to particle size because the competing hydrodynamic and acoustic forces scale differently with particle radius. The streaming-induced drag force scales linearly with particle size \citep{Wiklund2012,Hahn2015-zu}, whereas the acoustic radiation force scales with the particle volume \citep{Lenshof2012,Sujith2024}. This disparity gives rise to a characteristic crossover size that determines the dominant transport mechanism: particles larger than the crossover size are primarily driven by radiation forces, while smaller particles are advected by streaming flows. The combined influence of these mechanisms has been extensively studied in Newtonian fluids within rectangular microchannels \citep{ Hahn2015-zu}. For example, for polystyrene particles suspended in water at an excitation frequency of 2 MHz, the crossover size is $\approx2~\mu m$. Consequently, trapping particles below this size remains challenging due to the dominance of streaming-induced drag. Moreover, in the radiation-dominated regime, particles of different sizes but identical acoustic contrast factors often migrate to the same equilibrium locations, limiting size-selective trapping. These limitations motivate the search for alternate mechanisms to tune particle migration and trapping in acoustofluidic systems. 

Previous strategies for controlling acoustic particle trapping have largely relied on modifying device geometry \citep{Evander2012,Malik2022}, employing sharp-edge actuators to generate localized streaming flows \citep{Doinikov2020}, or using multiple transducers to tailor the acoustic field \citep{Evander2012}. While effective, these approaches often require complex device architectures and are not easily implemented. Moreover, the resulting trapping locations are typically fixed once the device geometry and actuation conditions are established, limiting dynamic tunability. These limitations motivate the search for alternative mechanisms that enable continuous control of trapping positions without altering channel geometry or acoustic actuation. Since particle migration is fundamentally governed by the competition between streaming-induced drag and acoustic radiation forces, tuning this balance, and consequently the crossover particle size, offers a potential route for manipulating particles across a broader size range. This raises the central question addressed in the present work: can fluid viscoelasticity be exploited to modulate this force balance, and thereby control particle trajectories and trapping locations?

Acoustic manipulation of particles in viscoelastic fluids has attracted growing interest due to its relevance in biological and complex-fluid systems~\citep{Doinikov2021a, Doinikov2021b,sujith2026fluidviscoelasticitycontrolsacoustic}. Many fluids encountered in acoustofluidic applications, including blood, plasma, sputum, saliva, synovial fluid and protein solutions, exhibit pronounced viscoelastic behaviour \citep{Thurston1972, Pan2009}. Accurately predicting particle migration in such media therefore requires understanding how fluid elasticity modifies the underlying acoustofluidic transport mechanisms. Previous studies have independently examined the influence of viscoelasticity on acoustic radiation forces \citep{Doinikov2021b, Sujith2024} and on acoustic streaming flows \citep{Doinikov2021a, Vargas2022, sujith2026fluidviscoelasticitycontrolsacoustic}. Radiation-driven migration has been shown to weaken under viscous effects but can be enhanced by fluid elasticity \citep{Sujith2024, Doinikov2021b}. Likewise, viscoelastic effects can suppress, enhance, or even reverse acoustic streaming flows \citep{sujith2026fluidviscoelasticitycontrolsacoustic, Doinikov2021a, Vargas2022}. These observations suggest that fluid rheology can substantially alter both acoustic radiation forces and streaming-induced drag, leading to potentially rich migration dynamics. However, despite these advances, the coupled influence of radiation forces and streaming drag on particle migration and trapping in viscoelastic fluids inside microchannels remain largely unexplored.

Acoustophoretic microfluidic devices commonly operate in a regime where the channel dimensions are comparable to the acoustic wavelength ($ \lambda_0$), such that the channel depth satisfies $D \sim \lambda_0$, the width is typically $W = \lambda_0/2$, and the viscous boundary layer remains much smaller than the channel dimensions ($\delta \ll D \sim W \sim \lambda_0$). Despite the practical importance of this configuration, a theoretical framework for particle trapping in viscoelastic fluids under these conditions is currently unavailable. While particle trajectories reveal the transient pathways governing the early stages of migration, they do not necessarily determine the final trapping state. Predicting migration and trapping behaviour therefore requires a combined analysis of both early-time migration and late-time equilibrium dynamics, which remains unexplored in viscoelastic acoustofluidic systems. In summary, previous studies have largely examined acoustic radiation forces and streaming flows independently in viscoelastic media, without addressing their coupled effects in realistic microchannel geometries. Consequently, how fluid rheology modifies the transition between streaming-dominated and radiation-dominated transport, and whether it can be exploited to tune particle trajectories and trapping locations in simple straight channels, remain open questions.

Here, we develop a unified theoretical and experimental framework to investigate particle migration and trapping in viscoelastic fluids subjected to ultrasound. The suspending fluid is modeled as an Oldroyd-B fluid \citep{oldroyd1950}, which captures the coupled effects of solvent viscosity, polymer elasticity, and fluid relaxation. The dynamics are governed by two key dimensionless parameters: the viscous diffusion number, $Dv$, which characterizes the relative contributions of solvent and polymer viscous transport, and the Deborah number, $De$, which compares the fluid relaxation timescale with the acoustic timescale. Using a perturbation analysis, we derive the oscillatory flow, boundary-layer dynamics, and steady streaming fields by extending classical acoustic streaming theories (\cite{Nyborg1965}, \cite{Lee1990}) to viscoelastic fluids (\cite{sujith2026fluidviscoelasticitycontrolsacoustic}). Acoustic radiation forces are incorporated through a viscoelastic framework (\cite{Sujith2024}), enabling prediction of particle motion under the combined influence of streaming-induced drag and radiation forces. Our formulation reveals a critical particle size that delineates the transition between streaming-dominated and radiation-dominated transport in viscoelastic fluids. The theoretical predictions are validated experimentally and demonstrate that fluid rheology can be used to continuously tune particle trajectories and final trapping locations. The remainder of the paper is organized as follows. Section \S\ref{sec:2} presents the theoretical formulation for particle migration, trapping and critical particle size, \S\ref{sec:3} describes the experimental methodology, \S\ref{sec:4} examines early- and late-time trapping dynamics, and conclusions are presented in \S\ref{sec:6}.

\section{Theoretical model}\label{sec:2}

\begin{figure}
\centering
\includegraphics[clip,width=1\textwidth]{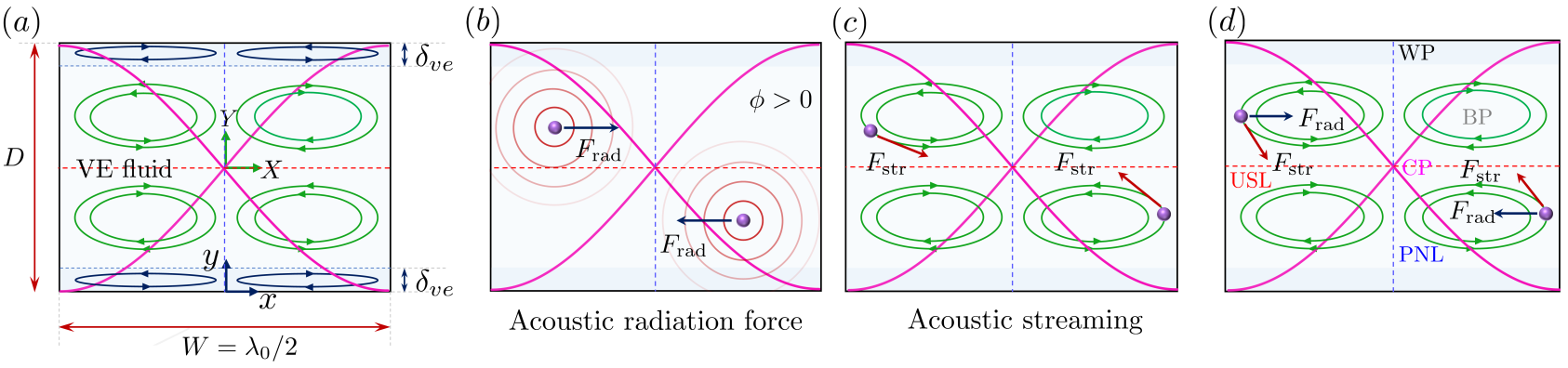}
\caption
{
\justifying{(a) Schematic of the  domain of study, corresponding to the cross section of the microchannel, where $W$ denotes the channel width and $D$ denotes the channel depth. The standing acoustic wave is along $X$ direction. Schematic illustrating the individual contributions of (b) acoustic radiation force $(F_{\mathrm{rad}})$, (c) acoustic streaming--induced drag force $(F_{\mathrm{str}})$, and their (d) combined effect on particle migration across the channel cross section. PNL denotes the pressure nodal line, USL the ultrasound symmetry line, CP the center point, and WP the wall point. Arrows indicate the directions of the respective forces.
 }
}
\label{fig:fig1}
\end{figure}

We consider the motion of a rigid spherical particle suspended in a viscoelastic fluid within a rectangular microchannel exposed to ultrasonic excitation, as illustrated in Fig.~\ref{fig:fig1}(a). The channel has width $W$ and depth $D$, and a Cartesian coordinate system is defined with its origin at the channel centre. The suspending fluid is modelled as an Oldroyd-B fluid \citep{oldroyd1950}, which captures the combined effects of viscous dissipation and elastic stress relaxation. In this model, the total dynamic viscosity is given by $\mu = \mu_p + \mu_s$, where $\mu_s$ and $\mu_p$ denote the solvent and polymer contributions, respectively, while fluid elasticity is characterised by the relaxation time $\tau$. The relative importance of viscous and elastic effects is governed by two dimensionless parameters. The first is the viscous diffusion number, $Dv=t_{d,s}/t_{d,p}=\nu_p/\nu_s$, which compares the solvent and polymer viscous diffusion timescales, where $t_{d,s}=(D/2)^2/(2\nu_s)$ and $t_{d,p}=(D/2)^2/(2\nu_p)$. Here, $\nu_s$ and $\nu_p$ denote the solvent and polymer contributions to the kinematic viscosity, respectively. The second is the Deborah number, $De=\tau/t_{ac}$, which compares the fluid relaxation timescale with the acoustic timescale $t_{ac}=1/\omega$, where $\omega$ is the angular frequency of acoustic excitation. Together, $Dv$ and $De$ provide a unified framework for characterising the coupled viscous-elastic response of fluids under acoustic forcing.

Initially, both the fluid and the suspended particle are at rest. Upon actuation, the piezoelectric transducer generates a one-dimensional standing acoustic wave along the $X$-direction, satisfying the half-wavelength resonance condition $W = \lambda_0/2$, where $\lambda_0$ denotes the acoustic wavelength. The imposed acoustic field induces first-order perturbations in the fluid pressure, density and velocity fields \citep{Doinikov2021a}. In the presence of a particle, scattering of the incident acoustic wave produces an acoustic radiation force that drives particle migration (see Fig.~\ref{fig:fig1}(b)). Simultaneously, viscous dissipation of acoustic energy within the fluid generates a steady second-order flow, commonly referred to as acoustic streaming, which exerts a hydrodynamic drag force on the particle (see Fig.~\ref{fig:fig1}(c)). The present study focuses on particle motion arising from the combined action of these competing mechanisms, as illustrated in Fig.~\ref{fig:fig1}(d). To systematically characterise particle migration and trapping, we define several reference locations within the channel. The standing wave establishes a pressure nodal line (PNL) at $X=0$ and an ultrasound symmetry line (USL) along $Y=0$. Their intersection defines the channel centre point (CP) at $(X,Y) = (0,0)$. The intersections of the PNL with the upper and lower channel walls are referred to as wall points (WP), while equilibrium locations within the fluid interior are termed bulk points (BP). These reference locations provide a convenient framework for describing particle trajectories and final trapping states throughout the study.

\subsection{Acoustic streaming in a viscoelastic fluid}\label{sec:2.1}

We first determine the acoustic streaming field generated within the viscoelastic fluid, as this flow governs the hydrodynamic drag acting on suspended particles. The fluid motion is described by the conservation of mass and momentum equations, which, together with the constitutive relation for an Oldroyd-B fluid, govern the acoustic response of the system. These equations are expressed as

\begin{equation}\label{eq:1}
\frac{\partial \rho}{\partial t}+\boldsymbol{\nabla} \cdot(\rho \boldsymbol{v})=0
\end{equation}
and
\begin{multline}\label{eq:2}
\rho\left(\frac{\partial \boldsymbol{v}}{\partial t}+\boldsymbol{v} \cdot \boldsymbol{\nabla} \boldsymbol{v}\right) =-\boldsymbol{\nabla} p+  \boldsymbol{\nabla} \cdot \mu_s\left[\boldsymbol{\nabla} \boldsymbol{v}+(\boldsymbol{\nabla} \boldsymbol{v})^T-\frac{2}{3}(\boldsymbol{\nabla} \cdot \boldsymbol{v}) \boldsymbol{I}\right]+\boldsymbol{\nabla} \cdot \boldsymbol{\sigma}.
\end{multline}
\noindent Here, $\boldsymbol{v}=v_x \boldsymbol{e}_x+v_y\boldsymbol{e}_y$ denotes the fluid velocity field, where $\boldsymbol{e}_x$ and $\boldsymbol{e}_y$ are unit vectors in the $x$ and $y$  directions, respectively, and $\mu_s$ is the solvent viscosity. In Eq. (\ref{eq:2}), $\boldsymbol{\sigma}$ represents the polymeric stress tensor associated with the Oldroyd-B model \citep{oldroyd1950}, given by 

\begin{equation}\label{eq:3}
\tau \tilde{\boldsymbol{\sigma}}+\boldsymbol{\sigma}=\mu_p\left[\boldsymbol{\nabla} \boldsymbol{v}+(\boldsymbol{\nabla} \boldsymbol{v})^T-\frac{2}{3}(\boldsymbol{\nabla} \cdot \boldsymbol{v}) \boldsymbol{I}\right].
\end{equation}

\noindent Here, $\mu_p$ denotes the polymer viscosity, $\tau$ is the fluid relaxation time, and $\tilde{\boldsymbol{\sigma}}$ represents the upper-convected derivative of the stress tensor ${\boldsymbol{\sigma}}$, defined as 
\begin{equation}\label{eq:4}
\tilde{\boldsymbol{\sigma}}=\frac{\partial \boldsymbol{\sigma}}{\partial t}+\boldsymbol{v} \cdot \boldsymbol{\nabla} \boldsymbol{\sigma}-\left(\boldsymbol{\nabla} \boldsymbol{v}\right)^T \cdot \boldsymbol{\sigma}-\boldsymbol{\sigma} \cdot \boldsymbol{\nabla} \boldsymbol{v}.
\end{equation}
Given $\tilde{\boldsymbol{\sigma}}$ represents the frame invariance of the stress tensor, where we introduce a convected time differentiation. We introduce perturbation theory to the above governing equations to get the fluctuations in the fluid field under an acoustic field.

To analyse the acoustically driven flow, we employ a regular perturbation expansion of the governing equations by decomposing the fluid density $\rho$, pressure $p$, velocity $\boldsymbol{v}$, and stress tensor $\boldsymbol{\sigma}$ into equilibrium, first-order, and second-order components \citep{Doinikov2021a,Sujith2024}. The equilibrium state is denoted by the subscript $0$, while subscripts $1$ and $2$ represent the first- and second-order perturbations, respectively. Accordingly, the fluid variables are expressed as 
\begin{equation}\label{eq:5}
\rho=\rho_0+\rho_1+\rho_2, \quad p=p_0+p_1+p_2
\end{equation}
\begin{equation}\label{eq:6}
\boldsymbol{v}=\boldsymbol{v}_1+\boldsymbol{v}_2, \quad \boldsymbol{\sigma}=\boldsymbol{\sigma}_1+\boldsymbol{\sigma}_2,
\end{equation}
where the viscoelastic fluid is assumed to be quiescent in the equilibrium state, such that $\boldsymbol{v}_0$ and $\boldsymbol{\sigma}_0$. Furthermore, the zeroth- and first-order fields are assumed to be known, allowing the second-order streaming flow to be determined subsequently. 

Substituting the perturbation expansions Eqs. (\ref{eq:5}) and (\ref{eq:6}) into (\ref{eq:1})-(\ref{eq:4}) and retaining only first-order terms, while neglecting higher-order nonlinear contributions, yields the first-order governing equations: 

\begin{equation}\label{eq:2.5}
\frac{\partial \rho_1}{\partial t}+\rho_0 \boldsymbol{\nabla} \cdot \boldsymbol{v}_1=0
\end{equation}
\begin{multline}\label{eq:2.6}
\rho_0 \frac{\partial \boldsymbol{v}_1}{\partial t}=-\boldsymbol{\nabla} p_1+  \boldsymbol{\nabla} \cdot \mu_s\left[\boldsymbol{\nabla} \boldsymbol{v}_1+\left(\boldsymbol{\nabla} \boldsymbol{v}_1\right)^T-\frac{2}{3}\left(\boldsymbol{\nabla} \cdot \boldsymbol{v}_1\right) \boldsymbol{I}\right]+\boldsymbol{\nabla} \cdot \boldsymbol{\sigma}_1
\end{multline}
and
\begin{equation}\label{eq:2.7}
\tau \frac{\partial \boldsymbol{\sigma}_1}{\partial \mathrm{t}}+\boldsymbol{\sigma}_1=\mu_p\left[\boldsymbol{\nabla} \boldsymbol{v}_1+\left(\boldsymbol{\nabla} \boldsymbol{v}_1\right)^T-\frac{2}{3}\left(\boldsymbol{\nabla} \cdot \boldsymbol{v}_1\right) \boldsymbol{I}\right].
\end{equation}
The first-order pressure and density perturbations are related through the equation of state, given by

\begin{equation}\label{eq:2.11}
p_1-p_0=\left(\rho_1-\rho_0\right) c_0^2.
\end{equation}
\noindent Here, $p_0$ and $\rho_0$ denote the equilibrium pressure and density of the fluid, respectively, and $c_0$ is the speed of sound in the fluid. Following \cite{Doinikov2021a,Vargas2022,Sujith2024}, we assume a harmonic time dependence of the first-order fields of the form $\mathscr{F}_1=\hat{\mathscr{F}}_1(r)e^{i \omega t}$, where $\hat{\mathscr{F}}_1$ denotes the spatial amplitude and $\omega=2\pi f$ is the angular frequency corresponding to the resonant frequency $(f)$ of the standing bulk acoustic wave. Substituting this form into Eqs. (\ref{eq:2.5})$-$(\ref{eq:2.7}) and simplifying yields 
\begin{equation}\label{eq:2.8}
i \omega \hat{\rho}_1+\rho_0 \boldsymbol{\nabla} \cdot \hat{\boldsymbol{v}}_1=0
\end{equation}
and
\begin{equation}\label{eq:2.9}
\rho_0 \frac{\partial \hat{\boldsymbol{v}}_1}{\partial t}=-\boldsymbol{\nabla} \hat{p}_1 +\boldsymbol{\nabla} \cdot \mu_c\left[\boldsymbol{\nabla} \hat{\boldsymbol{v}}_1+\left(\boldsymbol{\nabla} \hat{\boldsymbol{v}}_1\right)^T-\frac{2}{3}\left(\boldsymbol{\nabla} \cdot \hat{\boldsymbol{v}}_1\right) \boldsymbol{I}\right]
\end{equation}
\noindent where 
\begin{equation}\label{eq:014}
\mu_c=\mu_s+\frac{\mu_p}{1+i \omega \lambda}.
\end{equation}
\noindent Here, we introduce a complex viscosity $\mu_c$ to describe the effect of viscoelastic parameters on the first-order fluid field. The first-order acoustic fields do not contribute to the time-averaged flow \citep{Lenshof2012,Doinikov2021a}. In contrast, nonlinear interactions of the first-order fields generate a non-zero time-averaged second-order flow, which gives rise to acoustic streaming. Retaining terms up to second order in Eqs. (\ref{eq:1})-(\ref{eq:4}) and performing a time average over one acoustic cycle yields 

\begin{equation}\label{eq:016}
\rho_0 \boldsymbol{\nabla} \cdot\left\langle\boldsymbol{v}_2\right\rangle=-\boldsymbol{\nabla} \cdot\left\langle\rho_1 \boldsymbol{v}_1\right\rangle
\end{equation}
and
\begin{multline}\label{eq:017}
\rho_0\left\langle\boldsymbol{v}_1 \cdot \boldsymbol{\nabla} \boldsymbol{v}_1+\boldsymbol{v}_1 \boldsymbol{\nabla} \cdot \boldsymbol{v}_1\right\rangle=-\boldsymbol{\nabla}\left\langle p_2\right\rangle+\mu \boldsymbol{\nabla}^2\left\langle\boldsymbol{v}_2\right\rangle+\frac{\mu}{3} \boldsymbol{\nabla}\left(\boldsymbol{\nabla} \cdot\left\langle\boldsymbol{v}_2\right\rangle\right) -\tau \boldsymbol{\nabla} \cdot\left\langle\boldsymbol{v}_1 \cdot \boldsymbol{\nabla} \boldsymbol{\sigma}_1\right\rangle+\\ \tau \boldsymbol{\nabla} \cdot\left\langle\left(\boldsymbol{\nabla} \boldsymbol{v}_1\right)^T \cdot \boldsymbol{\sigma}_1\right\rangle  +  \tau \boldsymbol{\nabla} \cdot\left\langle\boldsymbol{\sigma}_1 \cdot \boldsymbol{\nabla} \boldsymbol{v}_1\right\rangle.
\end{multline}

\noindent The final three terms on the right-hand side of Eq. (\ref{eq:017}) arise from the Oldroyd-B constitutive model and represent additional body-force contributions associated with fluid viscoelasticity. The quantity $\left\langle \boldsymbol{v}_2 \right\rangle$ denotes the time-averaged second-order velocity field, corresponding to the acoustic streaming flow that must be determined within the microchannel. To determine the streaming flow within the microchannel, the effects of the lateral channel walls are incorporated by enforcing no-slip boundary conditions at $X=\pm W/2$ \citep{sujith2026fluidviscoelasticitycontrolsacoustic}. The resulting problem is solved using an iterative Fourier framework, in which the velocity field is expressed as an infinite series expansion. We introduce the non-dimensional coordinates $\tilde{X}=X/(W/2)$ and $\tilde{Y}=Y/(D/2)$, together with the channel aspect ratio $\alpha=D/W$. By matching the inner boundary-layer solution with the outer bulk flow, the final expressions for the streaming velocity components are obtained as \citep{sujith2026fluidviscoelasticitycontrolsacoustic}.
\begin{equation}\label{eq:2.063}
\langle v_{2,X}(\tilde{X},\tilde{Y})\rangle= C_s \ v_{\textrm{str}}\sum_{n=1}^{\infty}\Bigl\{C_{1n}\sin[n\pi\tilde{X}] A^{\parallel}(n\alpha,\tilde{Y})+C_{2n} A^\perp(n\alpha^{-1},\tilde{X}) \cos[n\pi\tilde{Y}]\Bigr\},
\end{equation}

\begin{equation}\label{eq:2.064}
\langle v_{2,Y}(\tilde{X},\tilde{Y})\rangle=C_s \ v_{\mathrm{str}}\sum_{n=1}^{\infty}\Bigl\{C_{1n}\cos[n\pi\tilde{X}] A^\perp(n\alpha,\tilde{Y})+C_{2n} A^\parallel(n\alpha^{-1},\tilde{X}) \sin[n \pi\tilde{Y}]\Bigr\}.
\end{equation}
Here, $C_s$ denotes the viscoelastic correction coefficient that quantifies the modification of acoustic streaming induced by fluid rheology, and is given as 
\begin{equation}\label{eq:2.45}
 C_s=\frac{2 \rho_0 c_0^2 I_1+2 De  Dv \mu_s^* I_2+ 2 De^2 I_3}{3(1+De^2)\chi^2 (\chi^2+\xi^2)^2 \rho_0 c_0^2  \mu_s^*(1+Dv)}. 
\end{equation}

\noindent In the above equation, $\mu_s^*$ is the dimensionless solvent viscosity, defined as \( \mu_s^* = \mu_s / \mu_f \), where \( \mu_f \) is the viscosity of the base Newtonian fluid used as a reference (Deionized water). $I_1,I_2$ and $I_3$ are functions of $Dv$ and $De$, and are defined as 
\begin{equation}\label{eq:2.46}
I_1=-4 \chi^4+9 \chi^2 \xi^2+\xi^4,
\end{equation}
\begin{equation}\label{eq:2.47}
I_2=2 \mu_f \omega \xi^2 (5 \chi^2+\xi^2)+\rho_0 c_0^2(\chi^2-\xi^2)(\chi^2+\xi^2)^2
\end{equation}
and
\begin{equation}\label{eq:2.48}
I_3=4 Dv \mu_s \omega \chi \xi (-\chi^2+\xi^2)+\rho_0 c_0^2 (-4 \chi^4+ 9 \chi^2 \xi^2+ \xi^4).
\end{equation}
Here, $\chi$ and $\xi$ are defined as
\begin{equation}\label{eq:5_26}
\chi=\sqrt{\sqrt{\Gamma^2+\Xi^2}+\Gamma},\ \quad
\xi=\sqrt{\sqrt{\Gamma^2+\Xi^2}-\Gamma},
\end{equation}
where $\Gamma$ and $\Xi$ are given by
\begin{equation}\label{eq:5_27}
    \Gamma=-\frac{1}{\mu_s^*}\Biggl\{\frac{Dv\ De}{Dv^{2}+2 Dv+(1+De^2)}\Biggr\}
\end{equation}
and
\begin{equation}\label{eq:5_28}
    \Xi=\frac{1}{\mu_s^*}\Biggl\{\frac{Dv+1+De^2 }{Dv^{2}+2 Dv+(1+De^2)}\Biggr\}.
\end{equation}

In Eqs. (\ref{eq:2.063}) and (\ref{eq:2.064}), the Rayleigh streaming velocity $v_{\textrm{str}}$ is defined as

\begin{equation}\label{eq:2.065}
    v_\textrm{str}=\frac{3}{8} \frac{(v_1^{d_0})^2}{c_0}.
\end{equation}
Here, $v_1^{d_0}$ denotes the amplitude of the first-order velocity field. The functions $A^\parallel(n\alpha,\tilde{Y})$, $A^\perp(n\alpha,\tilde{Y})$, $A^\perp(n\alpha^{-1},\tilde{X})$, and $A^\parallel(n\alpha^{-1},\tilde{X})$  account for geometric confinement effects associated with the rectangular channel and are defined as 

\begin{multline}\label{eq:2.066}
    A^\parallel(n\alpha,\tilde{Y})=\Lambda (n \alpha) \bigl\{ n \pi \alpha \cosh[n \pi \alpha] \cosh[n \pi \alpha \tilde{Y}]-\sinh[n \pi \alpha] \cosh[n \pi \alpha \tilde{Y}] \\ -n \pi \alpha \tilde{Y} \sinh[ n \pi \alpha] \sinh[n \pi \alpha \tilde{y}]\bigr\},
\end{multline}

\begin{multline}\label{eq:2.067}
    A^\perp(n\alpha,\tilde{Y})= \Lambda (n \alpha) \bigl\{ n \pi \alpha \cosh[n \pi \alpha] \sinh[n \pi \alpha \tilde{Y}]-\sinh[n \pi \alpha] \cosh[n \pi \alpha \tilde{Y}] \\ -n \pi \alpha \tilde{Y} \sinh[ n \pi \alpha] \cosh[n \pi \alpha \tilde{y}]\bigr\},
\end{multline}

\begin{equation}\label{eq:2.068}
    \Lambda(n \alpha)=\frac{1}{n \pi\alpha-\sinh[n\pi\alpha]\cosh[n\pi\alpha]}.
\end{equation}

The functions $A^\perp(n\alpha^{-1},\tilde{X})$, $A^\parallel(n\alpha^{-1},\tilde{X})$, and $\Lambda(n\alpha^{-1})$ are obtained from Eqs. (\ref{eq:2.066})--(\ref{eq:2.068}) through the substitutions $n\alpha \rightarrow n\alpha^{-1}$  \citep{sujith2026fluidviscoelasticitycontrolsacoustic}. The coefficients $C_{1n}$ and $C_{2n}$ are determined by enforcing the tangential no-slip boundary conditions at the channel walls \citep{Muller2013}, 

\begin{equation}
   \boldsymbol{C}_1 =[\mathbf{I}-\mathbf{A}^\perp(\alpha^{-1})\mathbf{A}^\perp(\alpha)]^{-1}\cdotp\mathbf{e}_1,
\end{equation}

\begin{equation}
   \boldsymbol{C}_2=-\mathbf{A}^\perp(\alpha)\cdot\boldsymbol{C}_1,
\end{equation}

\begin{equation}
    \mathbf{A}_{j,n}^\perp(\alpha^{-1})=(-1)^n\int_{-1}^1d\tilde{X} A^\perp(n\alpha^{-1},\tilde{X}) \sin(j\pi\tilde{X}).
\end{equation}

\subsection{Particle migration in acoustically driven viscoelastic fluids }\label{sec:2.3}

Under acoustic excitation, the motion of a particle suspended in a viscoelastic fluid is, in principle, influenced by acoustic radiation forces, hydrodynamic drag, inertial lift, and viscoelastic lift forces. In the absence of acoustic forcing, however, the fluid remains quiescent and no background flow is imposed in the present configuration. Consequently, inertial and viscoelastic lift effects are negligible under the operating conditions considered here \citep{Sujith2024, Doinikov2021b}. Particle migration is therefore governed primarily by the competition between the acoustic radiation force, ${F}_{\mathrm{rad}}$ and the drag force arising from acoustic streaming, ${F}_{\mathrm{drag}}$ whose relative magnitudes depend strongly on particle size\citep{Hahn2015-zu, Lenshof2012}. Balancing these two forces yields 
\begin{equation}\label{eq:2.072}
    \boldsymbol{F}_{\mathrm{drag}}= -\boldsymbol{F}_{\mathrm{rad}}.
\end{equation}

For a viscoelastic fluid exposed to half-wavelength acoustic resonance, the acoustic radiation force is given by (\cite{Sujith2024})

\begin{equation}\label{eq:2.073}
    \boldsymbol{F}_\mathrm{rad},{_X}=4\pi^2\Phi(\tilde{\kappa}_P,\tilde{\rho}_P,\tilde{\delta}_{ve},\tilde{\lambda}_{ve})\frac{a^3E_{ac}}W\sin\left(2\pi\frac XW+\pi\right) \boldsymbol{e}_X.
\end{equation}
Here, $a$ denotes the particle radius and $E_{ac}$ is the acoustic energy density. The parameters $\delta_{ve}$ and $\lambda_{ve}$  represent the acoustic boundary-layer thickness and viscous wavelength in the viscoelastic fluid, respectively \citep{Sujith2024}. The particle-to-fluid compressibility and density ratios are defined as $\tilde{\kappa}_P = \kappa_P / \kappa_0$ and $\tilde{\rho}_P = \rho_P / \rho_0$, respectively. The corresponding acoustic contrast factor for a viscoelastic fluid, $\Phi$, is given by 

\begin{equation}
\Phi\left(\tilde{\kappa}_{p}, \tilde{\rho}_P, \tilde{\delta}_{v e}, \tilde{\lambda}_{ve}\right)=\frac{1}{3} f_1(\tilde{\kappa}_P)+\frac{1}{2} f_2^r\left(\tilde{\rho}_P, \tilde{\delta}_{v e}, \tilde{\lambda}_{ve}\right),
\end{equation}
\begin{equation}
f_1(\tilde{\kappa}_{p})=1-\tilde{\kappa}_{p},
\end{equation}
\begin{equation}
f_2\left(\tilde{\rho}_P, \tilde{\delta}_{v e}, \tilde{\lambda}_{ve}\right)=\frac{2\left[1-\tilde{\gamma}\left(\tilde{\delta}_{v e}, \tilde{\lambda}_{ve}\right)\right](\tilde{\rho}_P-1)}{2 \tilde{\rho}_P+1-3 \tilde{\gamma}\left(\tilde{\delta}_{v e}, \tilde{\lambda}_{ve}\right)},
\end{equation}
\begin{equation}
\tilde{\gamma}\left(\tilde{\delta}_{v e}, \tilde{\lambda}_{ve}\right)=\frac{3 (1+i)\left[1+Q+i (1+P)\right]}{(P-i Q)^2}.
\end{equation}
In the above equations, $f_1$, and $f_2$ denote the monopole and dipole scattering coefficients, respectively. The parameters $P$ and $Q$ are given by
\begin{equation}\label{eq:2.8333}
P=\frac{1}{\tilde{\delta}_{v e}}+\frac{2 \pi}{\tilde{\lambda}_{ve}} \quad \textrm{and} \quad Q=\frac{1}{\tilde{\delta}_{v e}}-\frac{2 \pi}{\tilde{\lambda}_{ve}},
\end{equation}
where, $\tilde{\delta}_{ve}=\delta_{ve}/a$ and $\tilde{\lambda}_{ve}=\lambda_{ve}/a$.  If $\boldsymbol{u}_P$ denotes the acoustophoretic velocity of the particle and $\langle \boldsymbol{v}_2 \rangle$ represents the acoustic streaming velocity of the surrounding fluid, the Stokes drag force is given by 
\begin{equation}\label{eq:2.074}
    \boldsymbol{F}_{\mathrm{drag}}=6 \pi \mu a \chi_d [\langle \boldsymbol{v}_2 \rangle- \boldsymbol{u}_P].
\end{equation}
Here, $\chi_d$ denotes the drag correction coefficient that accounts for viscoelastic effects. 

Combining Eqs. (\ref{eq:2.072}), (\ref{eq:2.073}), and (\ref{eq:2.074}) yields the acoustophoretic velocity of the particle 
\begin{equation}\label{eq:2.080}
    \boldsymbol{u}_P=\frac{\boldsymbol{F}_\mathrm{rad}}{6 \pi \mu a \chi_d}+ \langle \boldsymbol{v}_2 \rangle.
\end{equation}
The particle velocity comprises contributions from acoustic radiation and acoustic streaming, such that $\boldsymbol{u}_P$$=$$\boldsymbol{u}_\mathrm{rad}$ + $\langle \boldsymbol{v}_2 \rangle$. Here, $\boldsymbol{u}_\mathrm{rad}=\boldsymbol{F}_\mathrm{rad}/6\pi \mu a \chi_d$ denotes the radiation-driven particle velocity. From Eq. (\ref{eq:2.073}), the acoustic radiation force acts solely along the $X$-direction, and consequently $\boldsymbol{u}_\mathrm{rad}$  possesses only an $X$-component. Substituting Eq. (\ref{eq:2.073}) into the above expression yields 
\begin{equation}\label{eq:2.081}
    \boldsymbol{u}_\mathrm{rad}=\frac{\Phi_{ve} \ \omega}{6 \nu \chi_d} \ \frac{4 E_{ac}}{\rho_0 c_0} \ a^2\sin\left(2\pi\frac XW+\pi\right) \boldsymbol{e}_X.
\end{equation}
Here, $\nu=\mu/\rho_0$ denotes the kinetic viscosity of the fluid. The particle velocity in the $X$-direction $(u_{P,_X})$ is governed by the combined effects of acoustic radiation and streaming. In contrast, since $F_\mathrm{rad}$ is strictly in the $X$-direction, the radiation-induced velocity vanishes along the $Y$-direction, and $u_{P,_Y}$ is determined solely by acoustic streaming. Substituting Eqs. (\ref{eq:2.063}-\ref{eq:2.065}) and (\ref{eq:2.081}) into (\ref{eq:2.080}), and simplifying, yields the dimensionless particle velocity components $(\tilde{u}_{P,_X}, \ \tilde{u}_{P,_Y})$ in the channel as 
\begin{multline}\label{eq:2_042}
    \tilde{u}_{P,_X}=\Biggl\{  \frac{\Phi_{ve}\omega}{6 \nu \chi_d \Xi} \ a^2  \ \sin\left(\pi \tilde{X}   +  \pi\right) \\ +C_s  \ \sum_{n=1}^{\infty}\Bigl\{C_{1n}\sin[n\pi\tilde{X}] A^{\parallel}(n\alpha,\tilde{Y})+C_{2n} A^\perp(n\alpha^{-1},\tilde{X}) \cos[n\pi\tilde{Y}]\Biggr\},
\end{multline}

\begin{equation}\label{eq:2_043}
     \tilde{u}_{P,_Y}=C_s  \sum_{n=1}^{\infty}\Bigl\{C_{1n}\cos[n\pi\tilde{X}] A^\perp(n\alpha,\tilde{Y})+C_{2n} A^\parallel(n\alpha^{-1},\tilde{X}) \sin[n \pi\tilde{Y}]\Bigr\}.
\end{equation}
In the above equations, the dimensionless particle velocity $\tilde{u}_P=u_P/u_0$, where $u_0=(4 E_{ac} \Xi)/(\rho_0 c_0) $ is the characteristic acoustophoretic velocity scale and $\Xi=3/8$ \citep{Muller2013}. The dimensionless spatial coordinates are given by $\tilde{X}=X/(W/2),$ and $\tilde{Y}=Y/(D/2)$. In a rectangular channel, the vertical component of the acoustic streaming velocity vanishes along the central plane $\tilde{Y}=0$, referred to here as the ultrasound symmetry plane, such that $\tilde{u}_{p,_Y}=0$. Along this symmetry plane, the horizontal component of the particle velocity reduces to 
\begin{multline}
    \tilde{u}_{P,X} (\tilde{X},0)=\Biggl\{  \frac{\Phi_{ve}\omega}{6 \nu \chi_d \Xi} 
 \ \sin\left(\pi \tilde{X}   +  \pi\right) \\ +C_s \ \sum_{n=1}^{\infty}\Bigl\{C_{1n}\sin[n\pi\tilde{X}] A^{\parallel}(n\alpha,0)+C_{2n} A^\perp(n\alpha^{-1},\tilde{X}) \Biggr\}.
\end{multline}

\subsection{Critical particle size for acoustic trapping in viscoelastic fluids }

The critical particle size defines the transition between streaming-dominated and radiation-dominated transport and therefore provides a key criterion for controlling particle migration and trapping. To quantify this transition, we determine the critical particle radius, $(a_c)$, and the corresponding critical diameter, $(d_c = 2a_c)$, by equating the magnitudes of the acoustic radiation force $(F_\mathrm{rad})$ and the streaming-induced drag force $(F_\mathrm{drag})$ in a viscoelastic fluid. At the critical condition, the particle remains stationary $\boldsymbol{u}_P=0$, indicating an exact balance between these competing effects. Using Eqs. (\ref{eq:2.063}-\ref{eq:2.065}), (\ref{eq:2.081}), and (\ref{eq:2.080}), the critical particle radius in a viscoelastic fluid is obtained as 
\begin{equation}\label{eq:0_46}
    a_c=\sqrt{\frac{6 \nu}{\omega}\frac{\Xi}{\Phi_{ve}} \ \chi_d \ |C_s|}.
\end{equation}
For a Newtonian fluid, $\chi_d=1, \ C_s=1$, $\Phi_{ve}=\Phi(\tilde{\rho}_P,\tilde{\kappa}_P,\tilde{\delta})$, therefore Eq. (\ref{eq:0_46}) reduces to \cite{Muller2013}. 


Using Eq. (\ref{eq:0_46}), Eqs. (\ref{eq:2_042}) and (\ref{eq:2_043}) are simplified as
\begin{multline}\label{eq:2.083}
    \tilde{u}_{P,X}=\Biggl\{  \biggl(\frac{a}{a_c}\biggr)^2 |C_s|  \ \sin\left(\pi \tilde{X}   +  \pi\right) \\ +C_s  \ \sum_{n=1}^{\infty}\Bigl\{C_{1n}\sin[n\pi\tilde{X}] A^{\parallel}(n\alpha,\tilde{Y})+C_{2n} A^\perp(n\alpha^{-1},\tilde{X}) \cos[n\pi\tilde{Y}]\Biggr\},
\end{multline}

\begin{equation}\label{eq:2.084}
     \tilde{u}_{P,Y}=C_s  \sum_{n=1}^{\infty}\Bigl\{C_{1n}\cos[n\pi\tilde{X}] A^\perp(n\alpha,\tilde{Y})+C_{2n} A^\parallel(n\alpha^{-1},\tilde{X}) \sin[n \pi\tilde{Y}]\Bigr\}.
\end{equation}
Along the ultrasound symmetry plane, the components of the particle velocity reduce to

\begin{multline}
    \tilde{u}_{P,X} (\tilde{X},0)=\Biggl\{  \biggl(\frac{a}{a_c}\biggr)^2 |C_s| 
 \ \sin\left(\pi \tilde{X}   +  \pi\right) \\ +C_s \ \sum_{n=1}^{\infty}\Bigl\{C_{1n}\sin[n\pi\tilde{X}] A^{\parallel}(n\alpha,0)+C_{2n} A^\perp(n\alpha^{-1},\tilde{X}) \Biggr\},
\end{multline}

\begin{equation}
    \tilde{u}_{P,Y} (\tilde{X},0)=0.
\end{equation}
Equations (\ref{eq:2.083})  and (\ref{eq:2.084}) are solved numerically in Wolfram Mathematica using Lagrangian particle tracking to determine particle trajectories and the corresponding equilibrium trapping locations. Based on Eqs. (\ref{eq:2.083}), (\ref{eq:2.084}), and (\ref{eq:0_46}), particle migration in a viscoelastic fluid can be classified into three distinct regimes: (i) $a \gg a_c$, corresponding to radiation-dominated acoustophoresis; (ii) $a \ll a_c$, corresponding to streaming-dominated acoustophoresis; and (iii) $a \sim a_c$, where particle motion is governed by the competing influences of acoustic radiation and streaming. 

\section{Experimental}\label{sec:Experimental}\label{sec:3}

Experiments are performed using dilute suspensions of spherical fluorescent polystyrene particles of diameter 1 and 5 $\mu m$ (microParticles GmbH, Berlin) dispersed in viscoelastic fluids. The test fluids consist of aqueous solutions of polyethylene oxide (PEO) and methyl cellulose (MC) (Sigma-Aldrich), prepared by dissolving the polymer powders in deionized water at prescribed weight concentrations. The solutions are gently stirred under controlled heating to ensure complete dissolution and are subsequently allowed to equilibrate at room temperature for several days to achieve full hydration and stable rheological properties.

The rheological properties of the test fluids are characterised using an Anton Paar rheometer, supplemented by values reported in the literature \citep{Wu2023, Wu2024, Milliken1990, sujith2026fluidviscoelasticitycontrolsacoustic, Morozova2018, Bykurganc2023, Nakagawa2024}. For PEO (1 MDa) at $C \approx 0.10\%$, the measured viscosity is $\mu \approx 1.69~\mathrm{mPa\,s}$ and the relaxation time is $\tau \approx 0.14~\mu s$, corresponding to $Dv \approx 0.9$ and $De \approx 1.75$. For MC at $C \approx 0.3\%$, the viscosity is $\mu \approx 6.23~\mathrm{mPa\,s}$ with $Dv \approx 6$, while the relaxation time lies in the range $\tau \approx O(10^{-4} - 10^{-3})~\mathrm{s}$, leading to $De \approx O(10^2 - 10^3)$ \citep{Morozova2018, Bykurganc2023, Nakagawa2024}. It is noted that the exact value of the relaxation time of MC solution is not critical, as the present experiments are designed to provide a qualitative understanding of particle migration and trapping behavior. Therefore, the observed trends are not sensitive to moderate variations in the exact value of $\tau$ or $De$ of MC solution. Furthermore, deionized water is considered as the base solvent in both experiments and analysis.

\begin{figure}
\centering
\includegraphics[clip,width=1\textwidth]{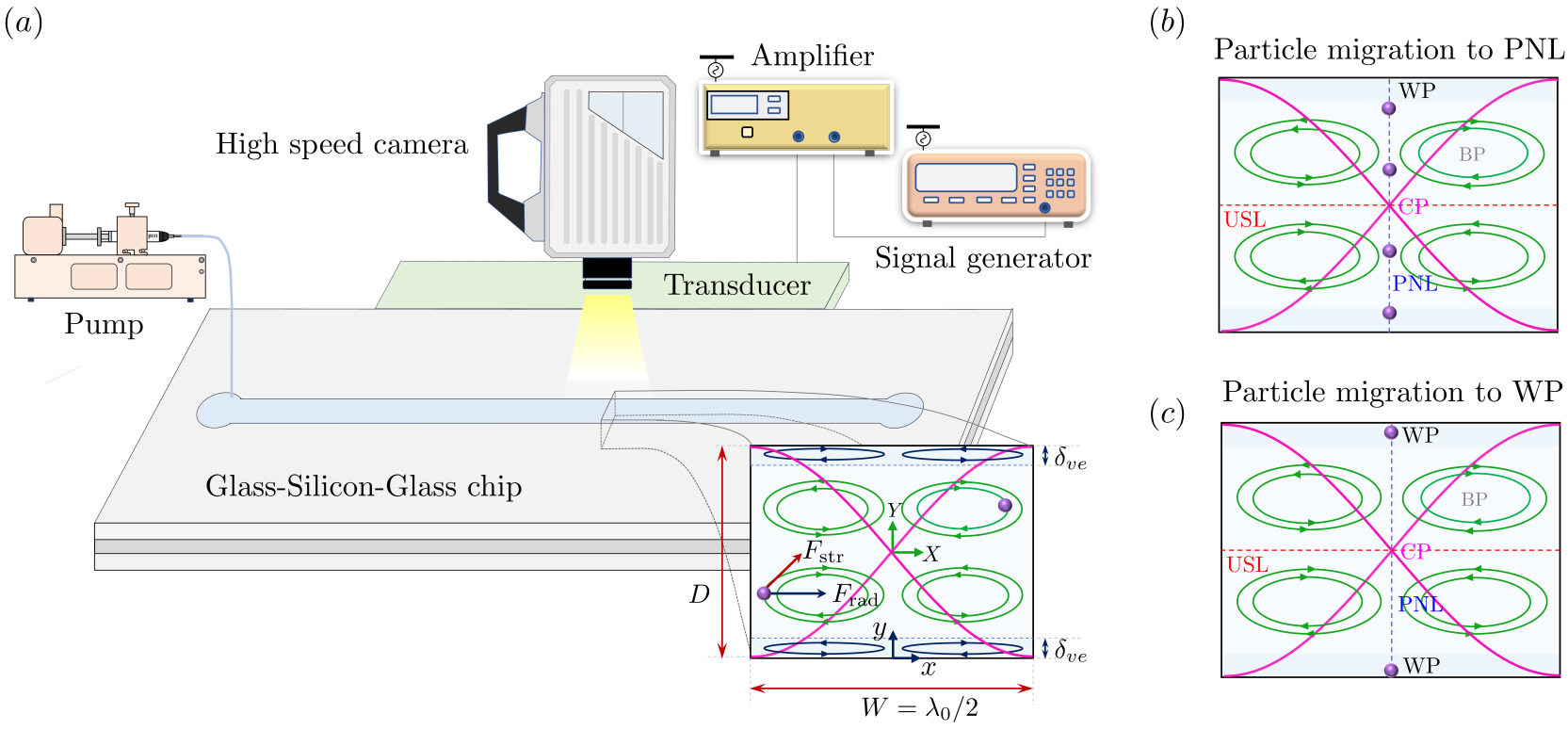}
\caption
{
\justifying{(a) Schematic of the experimental setup. Inset shows the cross section of the microchannel, where $W$ denotes the channel width and $D$ denotes the channel depth. A standing acoustic half-wave is established along the width-direction. Schematic illustrating the final equilibrium positions of particles under the combined effect of radiation force and streaming-induced drag in a radiation dominated regime: (b) Arrangement of particles at the pressure nodal line and (c) Trapping of particle at the wall point. PNL denotes the pressure nodal line, USL the ultrasound symmetry line, CP the common point, and WP the wall point. Arrows indicate the directions of the respective forces. }
}
\label{fig:fig02}
\end{figure}

The experiments are conducted using a glass–silicon–glass microfluidic device, shown in Fig.~\ref{fig:fig02}(a). The device is fabricated using standard photolithography followed by deep reactive ion etching (DRIE). The microchannel pattern is first transferred onto a 3-inch, $\langle 100 \rangle$-oriented silicon wafer of thickness 300 $\mu m$ using a positive photoresist (MICROPOSIT S1813) and ultraviolet lithography. The patterned silicon is subsequently etched to form a rectangular microchannel of length 20 mm, width 400 $\mu m$, and depth 300 $\mu m$. The channel is then sealed by anodically bonding borosilicate glass substrates of thickness 500 $\mu m$ to both sides of the silicon wafer at 450 $^\circ$C  under an applied voltage of 1000 V. Finally, inlet and outlet ports are created through the glass layers to enable fluid delivery. 

The viscoelastic suspension is introduced into the microchannel using a high-precision syringe pump (Cetoni GmbH, neMESYS). Acoustic excitation is generated by bonding a 2.0 MHz piezoelectric transducer (Sparkler Ceramics) to the bottom glass substrate using epoxy adhesive. The transducer is driven by a sinusoidal signal from a function generator (Rohde \& Schwarz, SMC100A) and amplified using a broadband power amplifier (Amplifier Research, 75A250A), with an operating power range of 10–1000 mW. The resonance frequency of the device is determined experimentally by introducing a suspension of 15 $\mu m$ polystyrene particles and sweeping the excitation frequency from 1.85 to 2.1 MHz. A resonance condition is observed at 1.93 MHz, at which the migration and trapping dynamics of 1 and 5 $\mu m$ particles are recorded using a high-speed camera (Phantom, V2512). Under acoustic excitation, particles migrate within the channel cross-section toward distinct equilibrium locations, such as the pressure nodal line (PNL) or, in some cases, the wall points (WP), as illustrated in Fig.~\ref{fig:fig02}(b) and (c), respectively. To accurately resolve particle trapping locations throughout the channel depth, confocal fluorescence microscopy (Olympus, IX83) is employed. Optical sectioning is performed by acquiring image stacks at multiple focal planes across the channel depth, which are subsequently reconstructed to obtain the three-dimensional distribution of trapped particles. This approach enables precise identification of equilibrium trapping locations within the microchannel.

\section{Results and discussion}\label{sec:4}

We investigate the coupled effects of acoustic streaming and acoustic radiation forces on particle migration and trapping in a viscoelastic-fluid-filled microchannel. In Newtonian fluids, when acoustic radiation forces dominate, particles with a positive acoustic contrast factor migrate directly toward the pressure node along nearly rectilinear trajectories \citep{Lenshof2012}. In contrast, when acoustic streaming dominates, particle motion is governed by the hydrodynamic drag associated with steady streaming flows and follows the underlying vortical streamlines \citep{Wiklund2012}. In viscoelastic fluids, both the streaming-induced drag and acoustic radiation forces are altered by fluid rheology, leading to substantially richer migration dynamics. Their relative contributions are governed by the interplay of several characteristic time scales: the acoustic time scale, $t_{ac}=1/\omega$; the fluid relaxation time, $\tau$; the solvent viscous diffusion time, $t_{v,s}=(D/2)^{2}/(2\nu_s)$; and the polymer viscous diffusion time, $t_{v,p}=(D/2)^{2}/(2\nu_p)$. To characterise this competition, we define two dimensionless parameters: the Deborah number, $De=\tau/t_{ac}$, which quantifies elastic effects, and the viscous diffusion number, $Dv=t_{v,s}/t_{v,p}$, which characterises the relative contributions of solvent and polymer viscosity. These competing timescales ultimately determine particle trajectories and the resulting equilibrium trapping locations. To systematically characterise particle migration and trapping, we define several reference locations within the channel cross-section: the pressure nodal line (PNL), wall points (WP), centre point (CP), bulk points (BP), and the Ultrasound symmetry line (USL), as illustrated in Fig. \ref{fig:fig02}(b). These reference locations provide a convenient framework for distinguishing different migration pathways and identifying the corresponding equilibrium trapping states arising from the interplay of acoustic radiation and streaming.

\begin{figure}
\centering
\includegraphics[clip,width=1\textwidth]{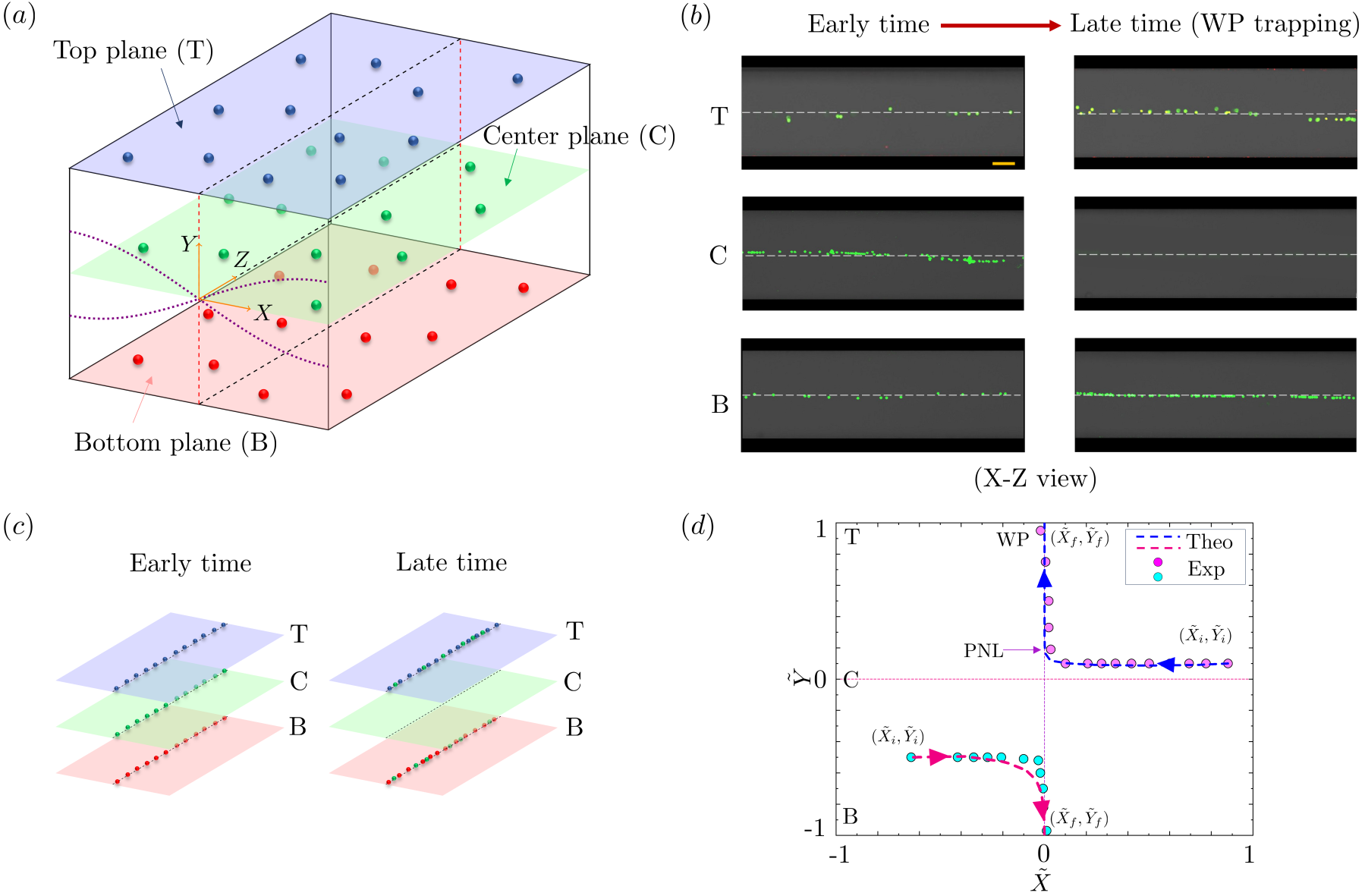}
\caption
{
\justifying{(a) Schematic of the microchannel illustrating the top (T), center (C), and bottom (B) planes, indicated in blue, green, and red, respectively, along with the corresponding particle locations at each plane. (b) Experimental images for 5 $\mu m$ polystyrene particles in a PEO solution with $Dv=0.9$ and $De=1.75$, depicting particle distributions at the top, center, and bottom planes in the early time ($\tilde{T}=3$) and late time ($\tilde{T}=6$) analyis in the WP trapping regime. The scale bar in the figure indicates 100 $\mu m$.  (c) Schematic of particle patterning at different planes during early and late time analysis of wall point (WP) trapping. 
(d) Comparison between the experimental and theoretical particle trajectories corresponding to wall point (WP) trapping for the experimental condition in (b). }
}
\label{fig:Fig_7_3}
\end{figure}

In  \S\ref{sec:4.01}, we first present experimental observations highlighting the coupled effects of acoustic radiation and streaming-induced drag on particle migration. Motivated by these observations, \S\ref{sec:4.1} employs theoretical analysis to explore a broader parameter space and systematically isolate the roles of viscous and elastic effects in shaping particle trajectories. In \S\ref{sec:4.2}, we examine the underlying force balance governing migration by separately analysing the contributions of acoustic streaming and radiation forces. Section \S\ref{sec:4.3} then investigates the trapping dynamics through early- and late-time analyses of particle motion. Finally, \S\ref{sec:4.4} extends the framework to determine the critical particle size and examine how viscoelasticity modifies the transition between streaming- and radiation-dominated transport.

\subsection{Experimental observations of acoustic particle migration in viscoelastic fluids }\label{sec:4.01}

Particle trajectories provide insight into the migration pathways followed during the initial stages of motion. However, in several cases, particles undergo only limited displacement at early times, making it difficult to infer their eventual trajectories or final trapping locations from transient behaviour alone. To address this limitation, we perform both early- and late-time analyses by solving Eqs.~(\ref{eq:2.083}) and (\ref{eq:2.084}). The early-time analysis captures short-time migration trajectories, whereas the late-time analysis characterises the long-term evolution toward equilibrium trapping states. Together, these analyses provide a comprehensive description of the trapping dynamics arising from the coupled effects of acoustic radiation and streaming-induced drag in viscoelastic fluids.

The temporal evolution of particle motion is characterised using the dimensionless time $\tilde{T}=t/(L_c/u_0)$, where $L_c$ is the characteristic length scale, defined as $W/2$ along the $X$-direction and $D/2$ along the $Y$-direction, and $u_0 = (4E_{ac}\Xi)/(\rho_0 c_0)$ is the characteristic velocity scale. As discussed in \S \ref{sec:2}, both acoustic radiation and streaming-induced drag are strongly dependent on particle size. Accordingly, introducing the critical particle size, $a_c$, allows the migration behaviour to be classified into three regimes: (i) $a \gg a_c$, corresponding to radiation-dominated acoustophoresis; (ii) $a \ll a_c$, corresponding to streaming-dominated acoustophoresis; and (iii) $a \sim a_c$, where particle migration is governed by the competition between acoustic radiation and streaming-induced drag.

%
\begin{figure}
\centering
\includegraphics[clip,width=0.9\textwidth]{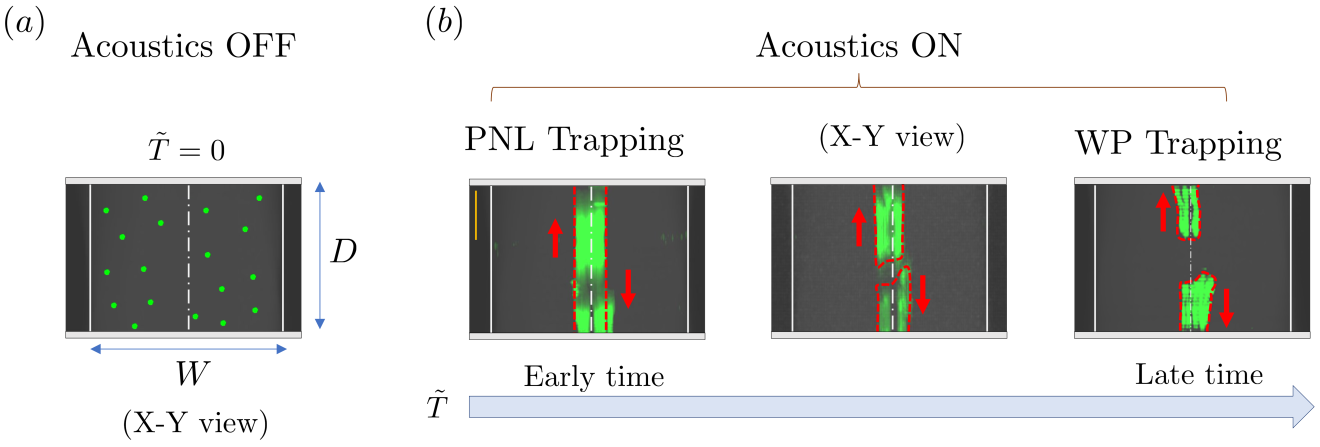}
\caption
{
\justifying{(a) Schematic representation of the distribution of 5 $\mu m$ polystyrene particles in a PEO solution ($Dv\approx0.9$ and $De\approx1.75$) across the channel cross-section, in the absence of the acoustic field. 
(b) Confocal microscopy images of the channel cross-section illustrating the migration of particles from the pressure nodal line (PNL) toward the wall points (WP) during early time $(\tilde{T}\approx 3)$ and late time $(\tilde{T}\approx 6)$ regimes. The scale bar in the figure indicates 100 $\mu m$.
 }
}
\label{fig:Fig_03}
\end{figure}

To validate the theoretical model, experiments are conducted in the radiation-dominated regime ($a > a_c$) to examine particle trapping dynamics in a viscoelastic fluid, as shown in Fig.~\ref{fig:Fig_7_3}. In these experiments, $5~\mu m$ polystyrene particles are suspended in a PEO solution characterised by $Dv\approx0.9$ and $De\approx1.75$, for which the acoustic contrast factor is positive ($\Phi>0$). Prior to acoustic actuation, the particles are uniformly distributed across the channel cross-section, as shown in Fig.~\ref{fig:Fig_7_3}(a). To resolve the three-dimensional trapping dynamics, particle distributions are recorded at three focal planes: top, centre, and bottom, each represented by a distinct colour (see Fig.~\ref{fig:Fig_7_3}(a)). Experimental snapshots are shown in Fig.~\ref{fig:Fig_7_3}(b). At early times ($\tilde{T}=3$), particles accumulate along the pressure nodal line (PNL) at $X=0$, indicating nodal-line trapping (Fig.~\ref{fig:Fig_7_3}(c,d)). At later times ($\tilde{T}=6$), particles migrate away from the nodal line toward the upper and lower walls, resulting in wall-point (WP) trapping. This two-stage migration behaviour arises from the sequential dominance of different forces. Immediately after actuation, the standing acoustic wave generates a strong radiation force that rapidly drives particles toward the pressure node. Once the particles reach the nodal line, the radiation force vanishes, and their subsequent motion is governed by streaming-induced drag, which transports them toward the channel walls. The theoretical predictions capture this transition and show good agreement with the experimental observations (see Fig.~\ref{fig:Fig_7_3}(d)).

\begin{figure}
\centering
\includegraphics[clip,width=0.95\textwidth]{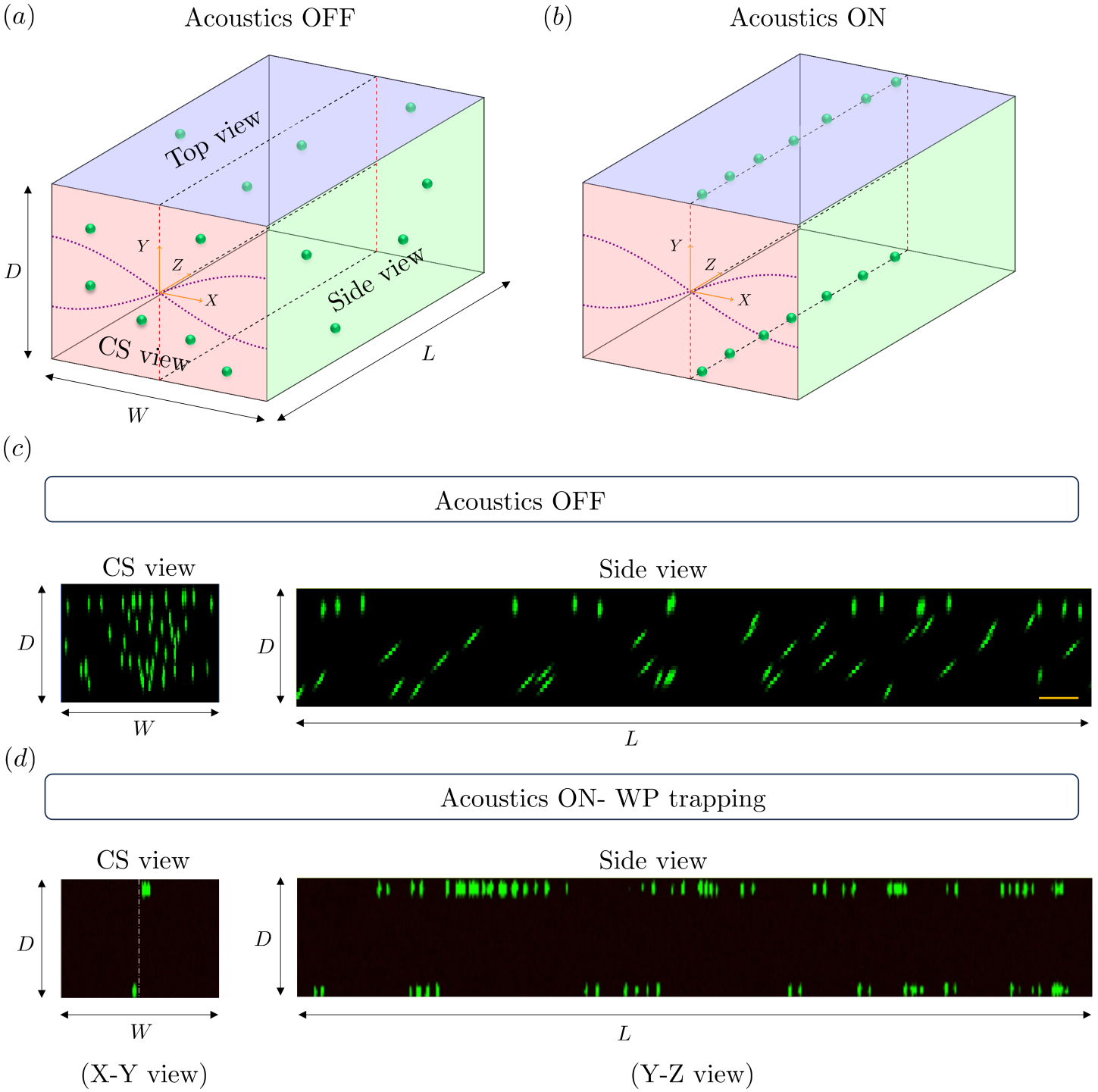}
\caption
{
\justifying{(a) Schematic of the microchannel illustrating the different perspectives used in the confocal experiments, including the cross-sectional (CS), top, and side views, with acoustic field OFF. (b) Schematic of wall-point (WP) trapping with acoustic field ON. Confocal microscopy images presenting the CS and side views of the channel with (c) acoustic field OFF, and (d) acoustic field ON, demonstrating wall-point (WP) trapping of 5 $\mu m$ polystyrene particles in a PEO solution ($Dv\approx0.9$ and $De\approx1.75$). The scale bar in the figure indicates 100 $\mu m$.
 }
}
\label{fig:Fig_7_5}
\end{figure}
\begin{figure}
\centering
\includegraphics[clip,width=0.95\textwidth]{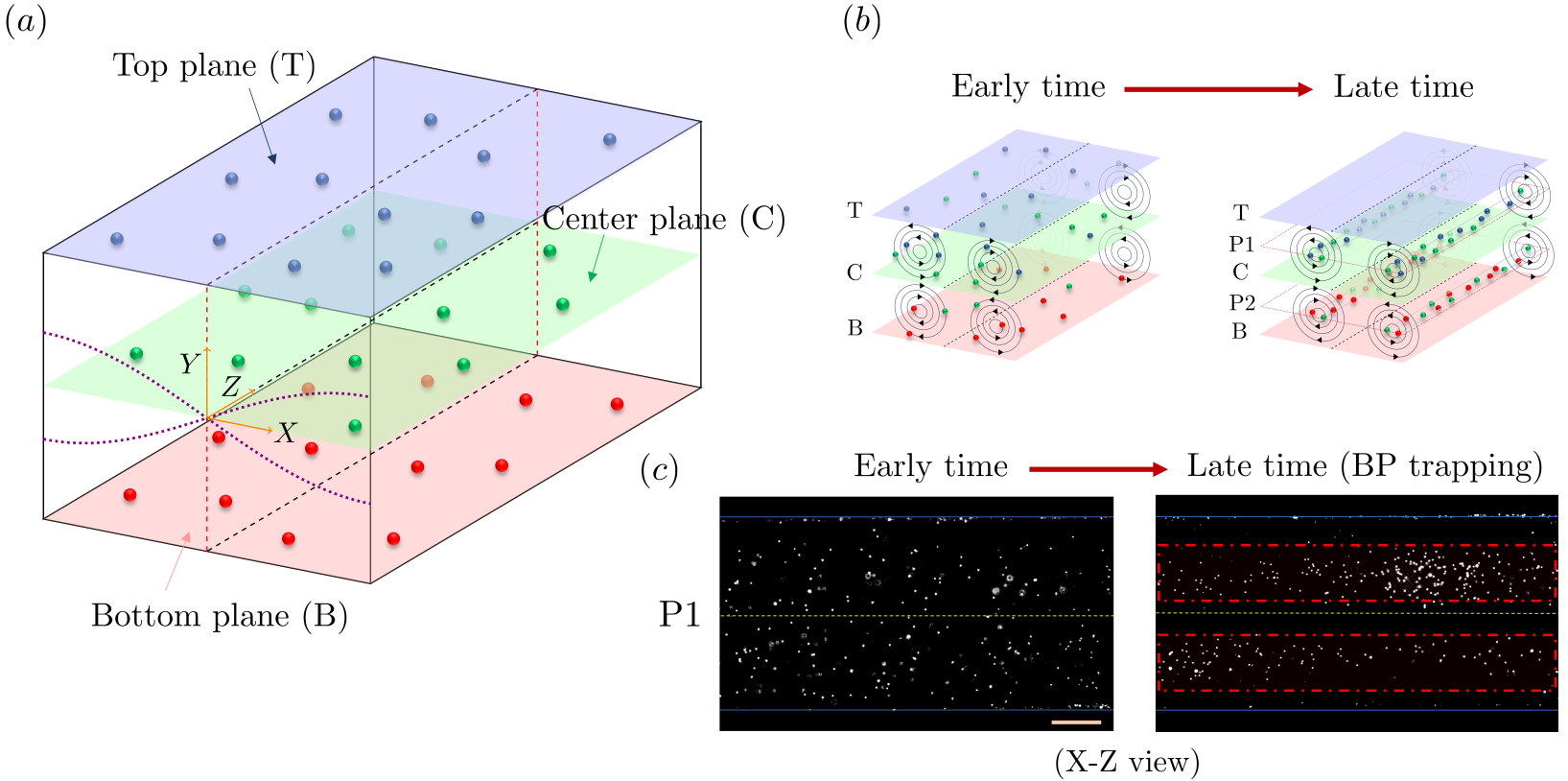}
\caption
{
\justifying{(a) Schematic of the microchannel illustrating the top (T), center (C), and bottom (B) planes, indicated in blue, green, and red, respectively, along with the corresponding particle locations at each plane. 
(b) Schematic representation of the particle patterning at different planes during the early and late time analysis of bulk point (BP) trapping. The streaming rolls are also shown. 
(c) Experimental images (top view) depicting distribution of 1 $\mu m$  polystyrene particles in a PEO solution ($Dv \approx 6,; De \sim O(100\text{$-$}1000)$) in the early and late time analyis, demonstrating bulk-point (BP) trapping across BP planes $P_1$ and $P_2$. The scale bar indicates 100 $\mu m$. 
 }
}
\label{fig:Fig_7_6}
\end{figure}
\begin{figure}
\includegraphics[clip,width=0.95\textwidth]{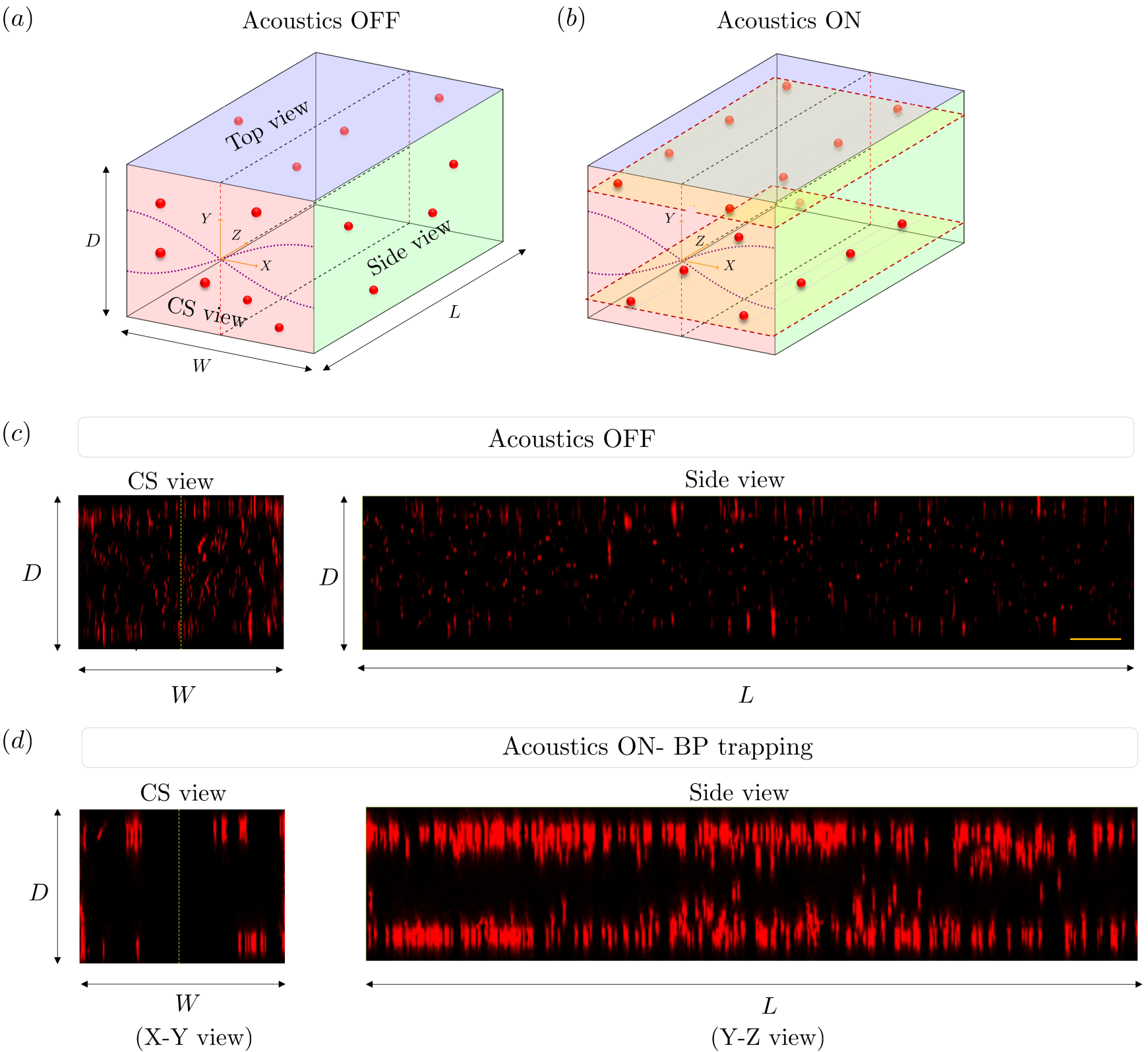}
\caption
{
\justifying{(a) Schematic of the microchannel illustrating the different perspectives used in the confocal experiments, including the cross-sectional (CS), top, and side views. (b) Schematic of bulk-point (BP) trapping with acoustic field ON. Confocal microscopy images presenting the CS and side views of the channel with (c) acoustic field OFF, and (d) acoustic field ON, demonstrating bulk-point (BP) trapping of 1 $\mu m$ polystyrene particles in a PEO solution ($Dv \approx 6,; De \sim O(100\text{$-$}1000)$). The scale bar in the figure indicates 100 $\mu m$.
 }
}
\label{fig:Fig_7_7}
\end{figure}

To resolve the full transition from pressure-nodal-line (PNL) trapping to wall-point (WP) trapping, confocal imaging is performed to examine the temporal evolution of particle distributions across the channel cross-section, as shown in Fig.~\ref{fig:Fig_03}. At early times, particles are concentrated primarily along the PNL. As time progresses, this trapped particle band progressively splits about the channel centre, with particles migrating toward the upper and lower walls. At late times, the particles reach the channel boundaries, consistent with the trapping behaviour presented in Fig.~\ref{fig:Fig_7_3}. The resulting wall-trapping distribution along the channel length is further examined using confocal cross-sectional and side-view images shown in Fig.~\ref{fig:Fig_7_5}. In the absence of acoustic excitation, particles remain uniformly distributed throughout the channel cross-section and along its length. Following acoustic actuation in the late-time regime, particles become localised predominantly near the upper and lower channel walls at $X=0$. The side-view images further confirm the depletion of particles near the channel centre and their accumulation near the channel boundaries.

In contrast to the radiation-dominated regime, particle motion becomes streaming-dominated when the particle size satisfies $a < a_c$, such that the streaming-induced drag force exceeds the acoustic radiation force. To experimentally probe this regime, $1~\mu m$ polystyrene particles are suspended in a methyl cellulose solution: $C \approx 0.3 \%,, Dv \approx 6,; De \sim O(100\text{$-$}1000)$ 
\citep{Nakagawa2024,Bykurganc2023}. The temporal evolution of particle distributions from the initial state (see Fig.~\ref{fig:Fig_7_6}(a)) to early ($\tilde{T}=3$) and late times ($\tilde{T}=10$) are shown schematically and experimentally in Fig.~\ref{fig:Fig_7_6} (b) and (c). 

Initially, particles are nearly uniformly distributed across the channel cross-section. Following acoustic actuation, they are transported along the streaming vortices and progressively migrate toward the vortex-eye regions. At later times, the particle distribution becomes increasingly localised near the vortex-eye plane, indicating bulk-point (BP) trapping (see Fig.~\ref{fig:Fig_7_6}(b) and (c)). This behaviour arises because acoustic radiation forces are negligible in this regime, and particle motion is governed primarily by acoustic streaming. Since the fluid remains quiescent in the absence of acoustic forcing, the resulting trapping behaviour is also sensitive to external perturbations, and even weak flow disturbances can influence the observed particle distribution. To further verify bulk trapping, confocal microscopy experiments are performed, as shown in Fig.~\ref{fig:Fig_7_7}. Prior to acoustic excitation, particles remain uniformly distributed throughout the channel cross-section and along its length (see Fig.~\ref{fig:Fig_7_7}(a) and (c)). In the late-time regime, particles accumulate at discrete locations within the channel, providing clear evidence of BP trapping (see Fig.~\ref{fig:Fig_7_7}(b) and (d)). Taken together with the radiation-dominated experiments discussed earlier, these observations demonstrate that particle migration in viscoelastic acoustofluidic systems is governed by the competition between acoustic radiation and streaming-induced drag. However, a deeper understanding of these migration pathways, and the possible emergence of additional trapping regimes under varying rheological conditions, requires a systematic analysis of the underlying force balance. To understand the role of viscoelasticity and elucidate the governing mechanisms, we turn next to theoretical modeling.

\begin{figure}
\centering
\includegraphics[clip,width=1\textwidth]{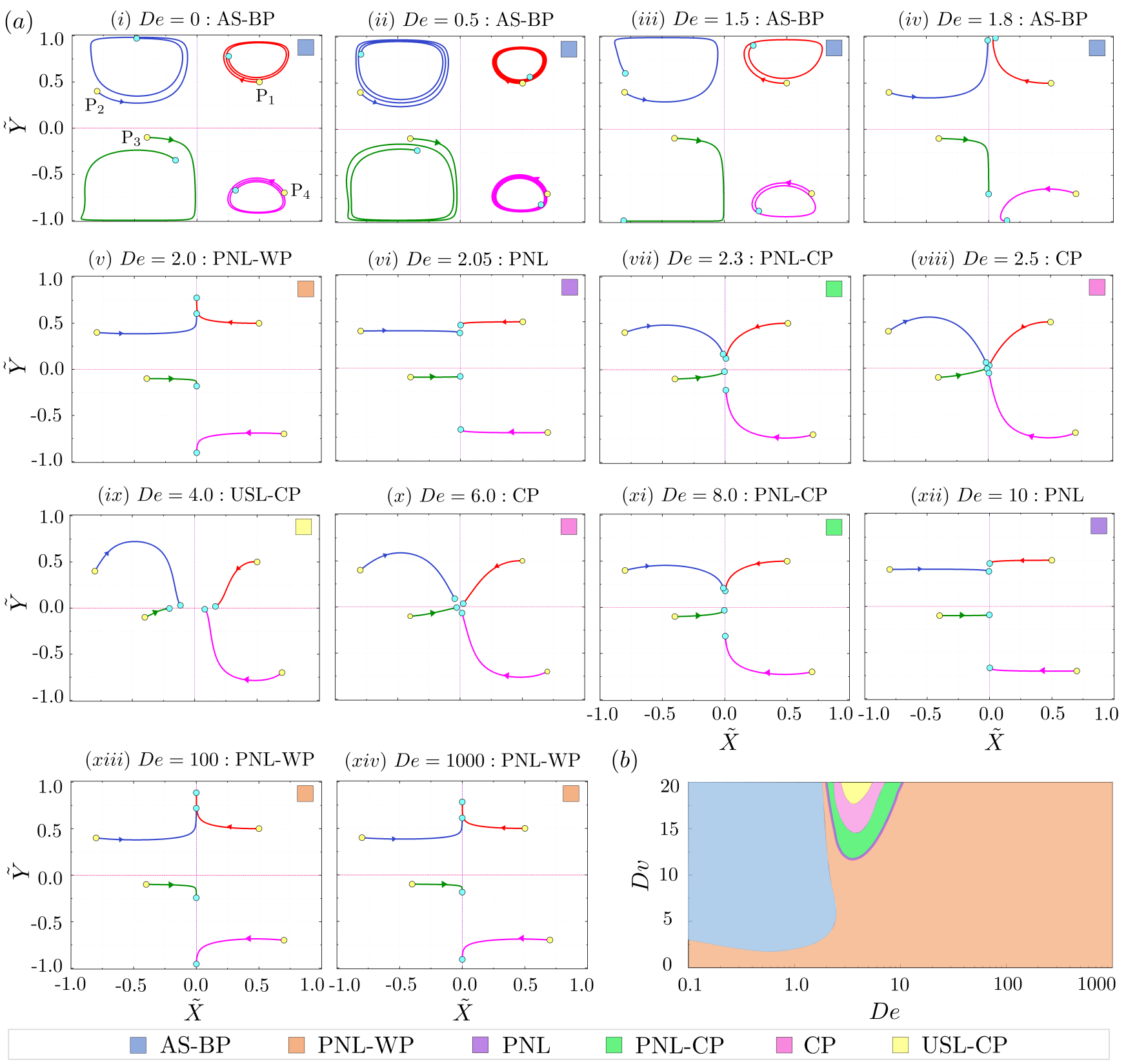}
\caption
{
\justifying{(a) Theoretically predicted trajectories of a $4 \ \mu$m polystyrene particle under the combined influence of acoustic streaming and acoustic radiation force for varying Deborah numbers, $De=0,\ 0.5, \ 1.5, \ 1.8, \ 2, \ 2.05, \ 2.3, \ 2.5,\ 4, \ 6, \ 8, \ 10, \ 100$ and 1000, (marked by (i)-(xiv)) at a fixed viscous diffusion number $Dv = 20$. The initial particle position at $\tilde{T} = 0$ is shown in yellow (see points $P_1-P_4$ in (a)-i), and the final position at $\tilde{T} = 10$ is shown in light cyan. (b) Regime map showing the six different trapping regimes: Acoustic streaming dominated bulk-point trapping (AS-BP), Pressure nodal line to wall-point trapping (PNL-WP), Pressure nodal line trapping (PNL), Pressure nodal line to centre-point trapping (PNL-CP), Centre-point trapping (CP), and Ultrasound symmetry line to centre-point trapping (USL-CP). 
 }
}
\label{fig:fig2}
\end{figure}

\subsection{Effect of viscoelasticity on particle trajectories under acoustic radiation and streaming}\label{sec:4.1}

Particle trajectories under ultrasound are governed by the competition between acoustic radiation and streaming-induced drag, as described by the theoretical model presented in \S\ref{sec:2.3}. According to Eqs.~(\ref{eq:2.073}) and (\ref{eq:2.074}), the radiation force scales with the cube of the particle diameter, whereas the streaming-induced drag varies linearly with particle size. This disparity in scaling makes the force balance highly sensitive to particle diameter, often limiting trajectory control to narrow size ranges when only these two mechanisms are considered. Consequently, achieving robust and tunable trapping across multiple spatial locations requires an additional mechanism to regulate particle motion. To address this limitation, we introduce viscoelasticity as a control parameter for tuning particle trajectories. The particle size is characterised by the blockage ratio, $\beta = d/(W/2)$, defined as the ratio of particle diameter to channel half-width. Fig.~\ref{fig:fig2} (a) illustrates the variation of particle trajectories with Deborah number ($De$) for $\beta = 0.02$ at $Dv = 20$. The polystyrene particle diameter is chosen as $d = 4~\mu$m, smaller than the critical size in the Newtonian limit ($De = 0$), $d_c \approx 9~\mu m$, as obtained from Eq.~(\ref{eq:0_46}). Since $d < d_c$, the system initially lies in the streaming-dominated regime for $De = 0$. Increasing $De$ progressively shifts the balance between acoustic radiation and streaming, enabling transitions across multiple migration regimes and allowing systematic exploration of the resulting trapping behaviour. 

Particles are initially released from four locations, $((\tilde{X}_i,\tilde{Y}_i)=$ $P_1(0.5,0.5)$, $ P_2(-0.8,0.4)$, $P_3(-0.4,-0.1)$, $P_4(0.7,-0.7))$, with one particle placed in each quadrant of the channel cross-section, as indicated by the yellow circles (see Fig.~\ref{fig:fig2} (a)). Particle trajectories are evaluated upto the dimensionless time $\tilde{T} = 10$, where $\tilde{T} = t/(L_c/u_0)$, and the corresponding final positions are marked by cyan circles. A Deborah number $De = 0$ corresponds to the purely viscous (Newtonian) limit, whereas $De > 0$ represents a viscoelastic fluid. For $De < 1$, viscous effects dominate the response, while elastic effects become increasingly important for $De >1$. In the Newtonian limit ($De = 0$), particles primarily follow the streaming vortices, resulting in nearly circular trajectories with no migration toward the pressure node (see Fig.~\ref{fig:fig2}~(a)-i). With a modest increase in elasticity ($De = 0.5$), the streaming rolls intensify, causing particles to move more rapidly along the closed trajectories. As $De$ increases further ($De >1$), this vortical motion is progressively suppressed. At $De = 1.5$, the streaming vortices become significantly weaker, while at $De = 1.8$, particle trajectories become nearly rectilinear and begin to align toward the pressure node, indicating the increasing dominance of acoustic radiation forces. We classify the regime spanning $0 \le De \lesssim 1.8$, in which particle motion is primarily governed by acoustic streaming and particles remain trapped within the bulk of the channel, as the acoustic streaming dominated bulk-point (AS–BP) trapping  regime (see Fig.~\ref{fig:fig2}~(a) i-iv).

When acoustic radiation forces overcome streaming-induced drag, particles migrate along nearly rectilinear paths toward the pressure node. At $De = 2$, particles reach the nodal line ($\tilde{X}=0$). In the vicinity of the pressure node, a weak transverse drift emerges, causing particles to migrate along the nodal line toward the ceiling or floor of the channel. This behaviour is classified as the pressure-nodal-line to wall-point (PNL–WP) trapping regime (see Fig.~\ref{fig:fig2}~(a)-v). For slightly higher elasticity ($De = 2.05$), this transverse drift vanishes and particles remain stably trapped at the pressure node, corresponding to the PNL regime (see Fig.~\ref{fig:fig2}~(a)-vi). With further increases in elasticity, additional migration pathways emerge. At $De = 2.3$, particles migrate from the nodal line toward the channel centre, defining the PNL–CP regime (see Fig.~\ref{fig:fig2}~(a)-vii). As $De$ increase further, particles follow fully curved trajectories toward the channel centre, corresponding to the CP regime, which is absent in Newtonian fluids (see Fig.~\ref{fig:fig2}~(a)-viii). At $De = 4$, particles first migrate toward the ultrasound symmetry line ($\tilde{Y}=0$) before moving toward the channel centre, resulting in the USL–CP regime (see Fig.~\ref{fig:fig2}~(a)-ix). Interestingly, with further increases in $De$, these regimes reappear in reverse order. Specifically, at $De = 6, 8, 10,$ and $100$, particles sequentially exhibit the CP, PNL–CP, PNL, and PNL–WP regimes (see Fig.~\ref{fig:fig2}~(a) x-xiii). At very high elasticity ($De = 1000$), the trajectory stabilises in the PNL–WP regime, and no further qualitative changes are observed (see Fig.~\ref{fig:fig2}~(a)-xiv). 

Overall, particle migration in viscoelastic fluids under the combined influence of acoustic radiation and streaming can be classified into six distinct regimes: AS–BP, PNL–WP, PNL, PNL–CP, CP, and USL–CP. The transitions between these regimes as functions of $De$ and $Dv$ are summarised in Fig.~\ref{fig:fig2}(b). At high viscous diffusion number ($Dv = 20$), all six regimes emerge at relatively low $De$, and except for the AS–BP regime, the remaining regimes recur periodically for $De \gtrsim 2$. At moderate $Dv$, only two regimes persist: AS–BP at low $De$ and PNL–WP at high $De$. At low $Dv$, only the AS–BP regime survives. These results demonstrate that tuning $De$ and $Dv$ provides a robust mechanism for systematically controlling particle trajectories and trapping states. In the following section, we examine the underlying force balance and velocity fields to elucidate the physical origin of these transitions. 

\begin{figure}
\centering
\includegraphics[clip,width=1\textwidth]{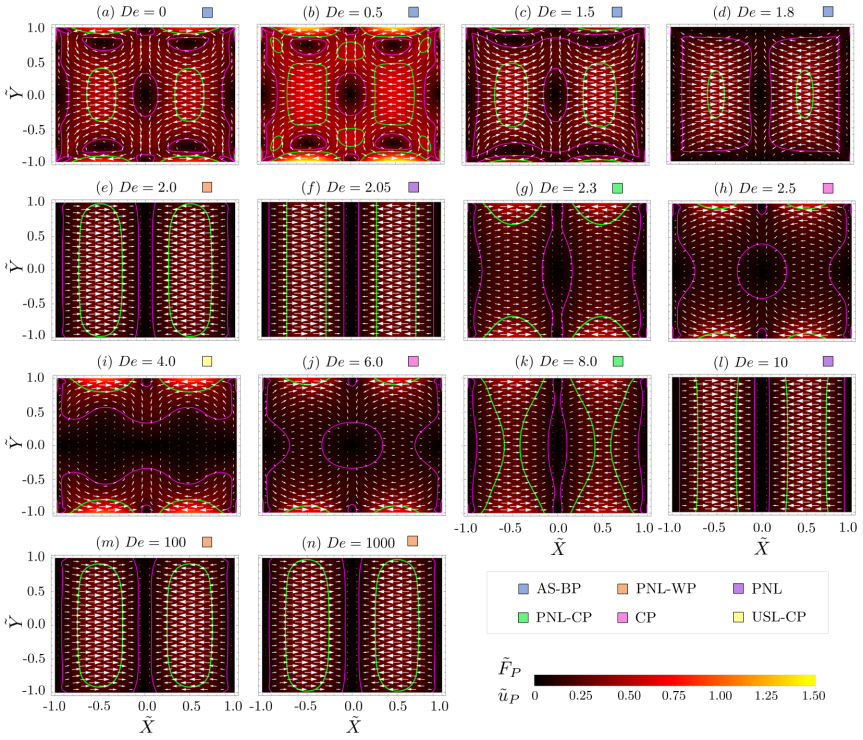}
\caption
{
\justifying{Theoretical results showing the variation of  dimensionless particle velocity $(\tilde{u}_P)$ or  dimensionless force $(\tilde{F}_P)$ variation across the cross section of the microchannel for different $De$ at $Dv=20$. The colour density indicate the magnitude of dimensionless particle velocity or force and the vectors indicate the direction. The magenta and green lines demarcate the region of very small and large particle velocity or force. }
}
\label{fig:fig3}
\end{figure}

\begin{figure}
\centering
\includegraphics[clip,width=1\textwidth]{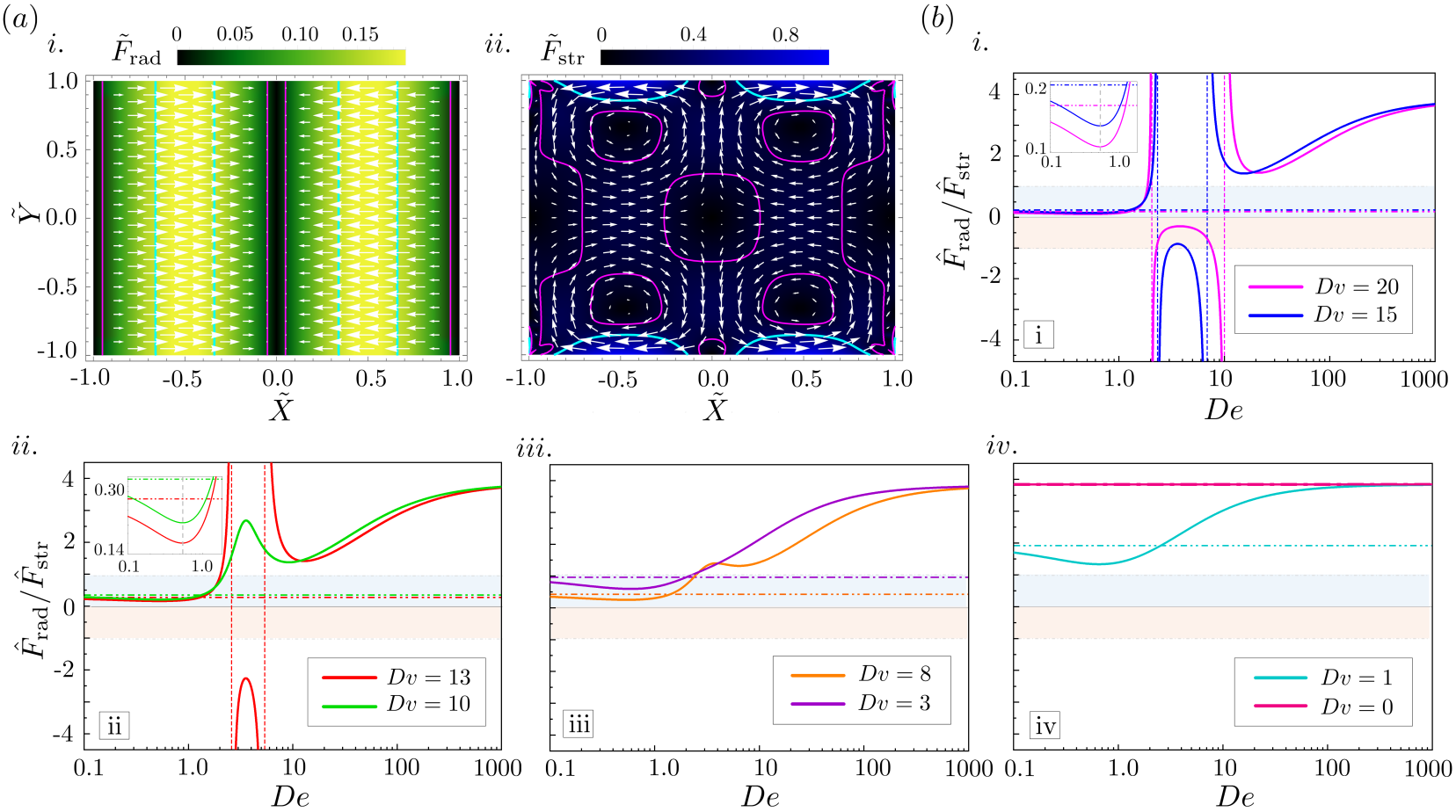}
\caption
{
\justifying{(a) Spatial variation of the dimensionless i. acoustic radiation force and ii. streaming-induced drag force across the channel cross-section. 
(b) Dependence of the ratio of dimensionless acoustic radiation force amplitude to streaming-induced drag force on the Deborah number $De$ for different values of the viscous diffusion number $Dv$ (i. $Dv$=20 and 15, ii. $Dv$=13 and 10, iii. $Dv$=8 and 3, iv. $Dv$=1 and 0).
The light blue and red shaded regions indicate regimes where streaming-induced drag dominates. The dash-dot lines indicate the value of $\hat{F}_\textrm{rad}/\hat{F}_\textrm{str}$ corresponding to $De=0$.
 }
}
\label{fig:fig04}
\end{figure}

\subsection{Effect of viscoelasticity on acoustic radiation and streaming forces}\label{sec:4.2} 

The particle trajectories shown in Fig.~\ref{fig:fig2} (a) are governed by the net force, $F_P$, resulting from the combined action of acoustic radiation and streaming-induced drag. This net force drives particle motion with velocity $u_P$. For convenience, the particle velocity is non-dimensionalised as $\tilde{u}_P = u_P/u_0$, and the corresponding force is expressed as $\tilde{F}_P = {6\pi \mu a u_P}/{F_0}$, where $F_0 = 6\pi \mu a u_0$. Under this definition, $\tilde{F}_P$ directly characterises the spatial variation of the normalised particle velocity. Fig. ~\ref{fig:fig3} shows the spatial distribution of $\tilde{F}_P$ (or equivalently $\tilde{u}_P$) across the channel cross-section for the same parameter values considered in Fig.~\ref{fig:fig2} (a). Regions of maximum force magnitude are highlighted in light green, while regions where the force becomes weak or nearly vanishes are indicated by magenta contours. In the purely viscous limit ($De = 0$), corresponding to the AS–BP regime in Fig.~\ref{fig:fig2} (a), the force magnitude is largest near the channel floor ($\tilde{Y}=-1$), ceiling ($\tilde{Y}=+1$), and along the symmetry plane ($\tilde{Y}=0$) at lateral positions $\tilde{X}\approx \pm 0.5$ (see Fig. ~\ref{fig:fig3}(a)). In contrast, regions of minimal force occur at the channel centre, near the sidewalls ($\tilde{X}=\pm 1$), along the floor and ceiling at $\tilde{X}=0$, and at four additional bulk locations corresponding to the vortex centres of the streaming rolls.

As elasticity increases (i.e. increasing $De$), the force magnitude initially increases and subsequently decreases, resulting in enhanced streaming at low $De$, followed by progressive suppression of streaming within the AS–BP regime (see Fig. ~\ref{fig:fig3} (a) - (d)). Notably, the locations of minimum force evolve significantly with increasing $De$. For $De>1$, the vortex-centre regions shift toward the floor and ceiling, while the central low-force region progressively extends toward these boundaries along the pressure nodal line. At $De = 1.8$, the central and vortex-centre regions merge, producing a minimum-force zone concentrated near the pressure node and adjacent to the channel walls (see Fig. ~\ref{fig:fig3}(d)). These results demonstrate that acoustic streaming can trap particles at distinct bulk locations, and that these trapping positions can be systematically tuned through the Deborah number. The progressive extension of the central low-force region also signals the increasing influence of acoustic radiation forces.

Beyond the AS–BP regime, the PNL–WP regime emerges at $De=2$ (see Fig.~\ref{fig:fig2} (a)). In this regime, the force vectors become nearly rectilinear; however, near the pressure node they deflect toward the floor and ceiling, forming a concave pattern (see Fig.~\ref{fig:fig3} (e)). This behaviour arises because the acoustic radiation force becomes negligible in the vicinity of the pressure node, allowing streaming-induced drag to transport particles toward the channel walls. At $De=2.05$, the streaming force near the pressure node is also strongly suppressed, and the minimum-force region collapses onto the nodal line (see Fig.~\ref{fig:fig3} (f)), resulting in stable trapping at the pressure node (PNL regime), consistent with Fig.~\ref{fig:fig2} (a).

With further increases in elasticity, the force field undergoes a qualitative reversal. At $De = 2.3$ (see Fig.~\ref{fig:fig3} (g)), the force vectors near the pressure node reorient toward the channel centre rather than toward the floor and ceiling, producing a convex minimum-force distribution. This indicates a reversal in the direction of the streaming-induced drag force, causing particles to migrate from the pressure node toward the channel centre, as observed in Fig.~\ref{fig:fig2}(a). At $De = 2.5$, the minimum-force region along the pressure nodal line disappears and contracts into a localised region at the channel centre, explaining the direct centreward migration shown in Fig.~\ref{fig:fig2} (a). At $De = 4$, this minimum-force region further elongates along the $X$-direction and aligns with the ultrasound symmetry line (see Fig.~\ref{fig:fig3} (i)), corresponding to the USL–CP regime shown in Fig.~\ref{fig:fig2} (a). At still higher $De$, these force distributions reappear in reverse sequence, explaining the recurrence of the CP, PNL–CP, PNL, and PNL–WP regimes observed in Fig.~\ref{fig:fig2} (a).

To decouple the individual contributions of acoustic radiation and streaming-induced drag, we examine their respective force fields in Fig.~\ref{fig:fig04}. The radiation force depends only on $\tilde{X}$ (see Eq.~\ref{eq:2.073}), whereas the streaming-induced drag force, $\boldsymbol{F}_\mathrm{str} = 6 \pi \mu a \langle \boldsymbol{v}_2 \rangle$, depends on both $\tilde{X}$ and $\tilde{Y}$ (see Eq.~\ref{eq:2.074}). Figure~\ref{fig:fig04}(a) shows the normalised radiation force, $\tilde{F}_{\mathrm{rad}} = F_{\mathrm{rad}}/F_0$, and the normalised streaming-induced drag force, $\tilde{F}_{\mathrm{str}} = F_{\mathrm{str}}/F_0$, across the channel cross-section for $De = 0$ and $Dv = 20$. To quantify their relative importance, Fig.~\ref{fig:fig04} (b) plots the ratio of the maximum radiation-force amplitude, $\hat{F}_{\mathrm{rad}}$, to the maximum streaming-drag amplitude, $\hat{F}_{\mathrm{str}}$, over a range of $De$ and $Dv$. The light-blue and red shaded regions correspond to streaming-dominated conditions ($|\hat{F}_{\mathrm{rad}}/\hat{F}_{\mathrm{str}}| < 1$), while the white region denotes radiation-dominated behaviour. For $Dv = 20$, this ratio initially decreases with increasing $De$, reflecting the strengthening of streaming-induced drag, and subsequently increases as streaming weakens. When $|\hat{F}_{\mathrm{rad}}/\hat{F}_{\mathrm{str}}| < 1$, the AS–BP regime prevails. As the ratio approaches unity, the PNL–WP regime emerges. A large positive ratio indicates negligible streaming drag, corresponding to stable PNL trapping. Negative values of the ratio signify a reversal in the direction of streaming-induced drag, giving rise to the PNL–CP and CP regimes. At higher $De$, the magnitude of this reversed drag decreases, leading to the re-emergence of the PNL–WP regime. For moderate viscous diffusion numbers ($Dv = 15$ and $Dv = 13$), similar transitions occur and all trapping regimes are recovered. In contrast, for lower values ($Dv = 10, 8,$ and $3$), the force ratio does not diverge because the streaming-induced drag never vanishes. As a result, the PNL, PNL–CP, and CP regimes disappear, and the dynamics remain confined to streaming- and radiation-dominated states. In the low-$Dv$ limit ($Dv = 1$ or $0$), acoustic radiation consistently dominates, leading exclusively to the PNL–WP regime.

These results demonstrate that the competition between acoustic radiation and streaming-induced drag determines the spatial distribution of minimum-force regions, which in turn governs particle trajectories and equilibrium trapping locations. The evolution of these regions with viscoelasticity further highlights the dynamic nature of the trapping process. Consequently, both early- and late-time dynamics must be considered to fully characterise particle migration across different regimes. The role of migration timescales in determining trapping behaviour is examined in the following section.

\begin{figure}
\centering
\includegraphics[clip,width=1\textwidth]{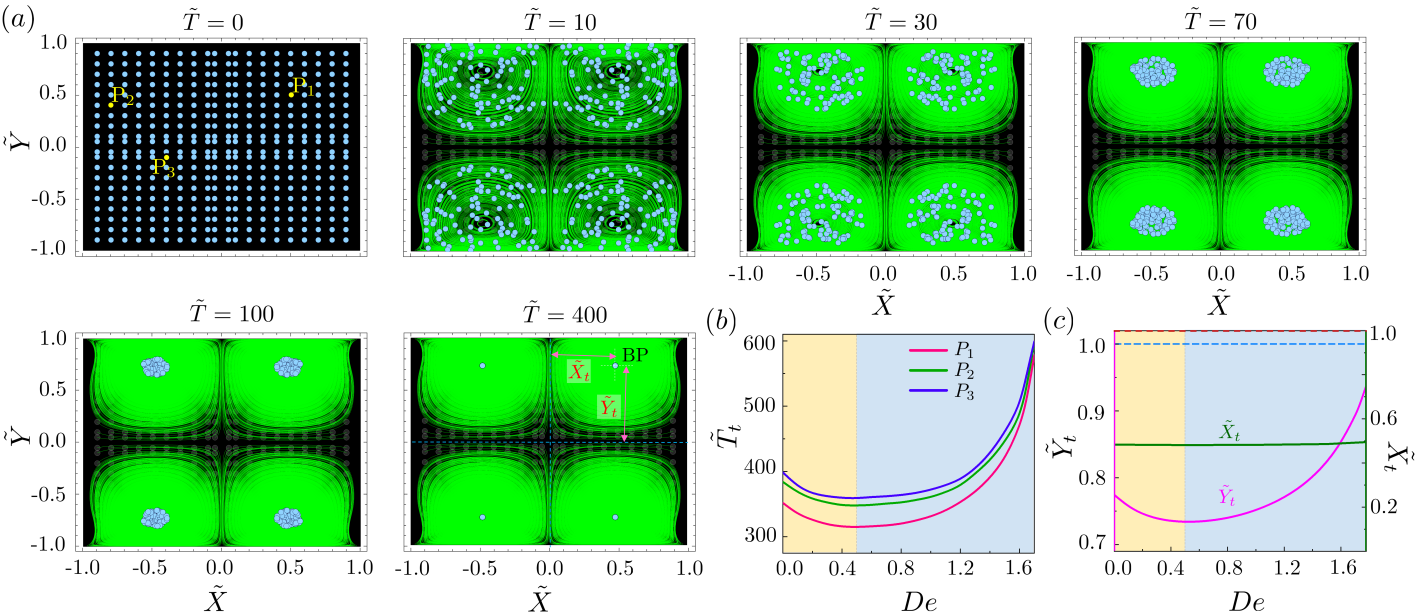}
\caption
{
\justifying{(a) Early- and late-time analysis of particle migration in the acoustic streaming dominated bulk-point (AS--BP) trapping regime at $De=0.5$ and $Dv=20$. 
A single particle is initialized at different locations across the channel cross-section. 
(b) Dependence of the trapping time, $\tilde{T}_t$, for three representative initial positions $P_1$, $P_2$, and $P_3$ 
(marked in (a)) on the Deborah number $De$ at $Dv=20$. 
The trapping coordinates are denoted by $(\tilde{X}_t,\tilde{Y}_t)$. 
(c) Variation of the trapping coordinates, $\tilde{Y}_t$ (magenta) and $\tilde{X}_t$ (green), with Deborah number $De$ at $Dv=20$. The blue dashed line indicates the channel ceiling, while the brown dashed line represents the right-side wall.
 }
}
\label{fig:fig05}
\end{figure}

\subsection{Early- and late-time acoustic particle migration and trapping in viscoelastic fluids }\label{sec:4.3} 

To examine the early- and late-time acoustic particle migration and trapping, we consider particles released from different initial locations across the channel cross-section, as illustrated in Fig.~\ref{fig:fig05} (a). The present analysis is restricted to single-particle dynamics, and inter-particle interactions are therefore neglected. Since both the trajectories and trapping locations are symmetric about the pressure nodal line (PNL) and the ultrasound symmetry line (USL), the analysis is confined to one quarter of the channel cross-section, which defines the region of interest. Below, we discuss each migration and trapping regime in detail.
\smallskip

\noindent \textit{Acoustic-streaming-dominated bulk point trapping (AS–BP) regime}: 
For the first regime identified in Fig.~\ref{fig:fig2}(b), namely the acoustic-streaming-dominated regime (AS–BP), the early- and late-time evolution of a particle for $De=0.5$ and $Dv=20$ is shown in Fig.~\ref{fig:fig05}. At early times, the particle follows the streaming rolls, circulating around the vortex centres under the action of the streaming-induced drag force. As time progresses, the particle gradually migrates toward the vortex centre. The weak but finite acoustic radiation force perturbs the streaming trajectories, enabling transitions between neighbouring vortices and facilitating migration toward the vortex centre at later times. Since the vortex centre corresponds to a location of minimal net force, it acts as an effective trapping point. As shown in Fig.~\ref{fig:fig05}(a), the final trapped position is attained at $\tilde{T}=400$. We denote the trapping location by $(\tilde{X}_t,\tilde{Y}_t)$ and the corresponding trapping time by $\tilde{T}_t$. The variation of the trapping time, $\tilde{T}_t$, with the Deborah number is shown in Fig.~\ref{fig:fig05}(b) for particles initially released from three different positions, $P_1(0.5,0.5)$, $ P_2(-0.8,0.4)$, $P_3(-0.4,-0.1)$ (marked in Fig.~\ref{fig:fig05} (a)). For all initial conditions, the trapping time exhibits a non-monotonic dependence on $De$, decreasing initially and subsequently increasing as $De$ increases. This behaviour can be interpreted in terms of the force ratio $\hat{F}_\mathrm{rad}/\hat{F}_\mathrm{str}$ shown in Fig.~\ref{fig:fig04}(b). At low $De$, the ratio decreases, reflecting an enhancement of the streaming-induced drag $\hat{F}_\mathrm{str}$, which accelerates particle migration and reduces the trapping time. At higher $De$, the ratio increases as the streaming drag weakens, leading to slower migration and, consequently, longer trapping times.

The AS–BP regime is observed throughout the range $0 \le De \lesssim 1.8$. Within this regime, the final trapping location varies systematically with $De$, and the dependence of the trapping position, $\tilde{Y}_t$ and $\tilde{X}_t$, on $De$ are shown in Fig.~\ref{fig:fig05}(c). In particular, the $\tilde{Y}$-coordinate of the vortex centre shifts with increasing $De$, while the $\tilde{X}$-coordinate remains nearly unchanged. This displacement of the vortex centre directly determines the corresponding variation in particle trapping location. As $De$ increases from zero to small finite values, $\tilde{Y}_t$ decreases, indicating that the trapping location moves toward the ultrasound symmetry line ($\tilde{Y}=0$). Beyond a critical value of $De$, however, the trend reverses and the trapping location shifts toward the channel ceiling. As $De$ approaches 1.8, the trapping position reaches $\tilde{Y}_t \approx 0.94$, i.e. close to the upper wall. These results demonstrate that, within the AS–BP regime, the late-time trapping location can be systematically tuned through $De$. 

This non-monotonic variation of the trapping location with $De$ arises from the corresponding non-monotonic behaviour of the force ratio $\hat{F}_\textrm{rad}/\hat{F}_\textrm{str}$ (see Fig.~\ref{fig:fig04}(b)). The trapping location corresponds to positions where the net force vanishes, i.e. where the acoustic radiation force and the streaming-induced drag are equal in magnitude and opposite in direction. At low $De$, the streaming-induced drag $\hat{F}_\textrm{str}$ increases with increasing elasticity (see \autoref{appA} Fig. \ref{fig:fig019}). Above the vortex centre, the radiation force and streaming drag oppose each other, with their magnitudes largest near the channel walls and decreasing toward the channel interior. As a result, an increase in $\hat{F}_\textrm{str}$ shifts the force balance away from the walls and toward the interior, causing the trapping location to move toward the ultrasound symmetry line. At higher $De$, however, the ratio $\hat{F}_\textrm{rad}/\hat{F}_\textrm{str}$ increases, reflecting a weakening of the streaming-induced drag. Under these conditions, the force balance can be achieved only closer to the channel boundaries, where the streaming drag remains sufficiently strong. Consequently, the trapping location shifts back toward the top and bottom walls. As $De$ approaches the upper limit of the AS–BP regime ($De \approx 1.8$), the trapping point lies very close to the channel boundary, signalling the transition to the next trapping regime.
\smallskip

\begin{figure}
\centering
\includegraphics[clip,width=1\textwidth]{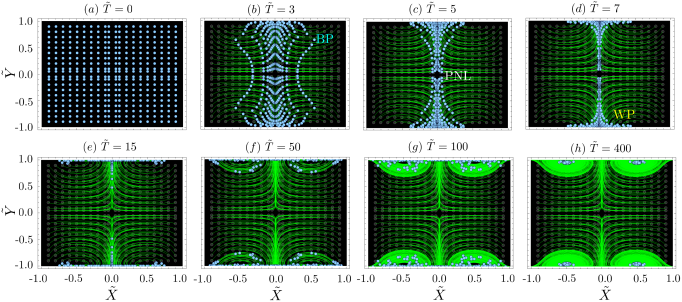}
\caption
{
\justifying{Early- and late-time analysis of particle migration at $De = 1.8$ and $Dv=20$, marking the transition from the AS--BP to the PNL--WP regime. A single particle is initialized at different positions across the channel cross-section.
 }
}
\label{fig:fig06}
\end{figure}

\noindent \textit{Pressure nodal line to wall point trapping regime (PNL–WP)}: We examine the early- and late-time dynamics of particle migration at $De=1.8$, corresponding to the transition between the AS–BP and PNL–WP regimes, as shown in Fig.~\ref{fig:fig06}. At early times, the particle migrates toward the pressure nodal line (PNL), following a curved trajectory forming a concave trapping pattern (see Fig.~\ref{fig:fig06}(c)). At intermediate times, the influence of streaming-induced drag becomes pronounced, redirecting the particle toward the ceiling or floor of the channel (see Fig.~\ref{fig:fig06}(d)). Subsequently, the particle moves along the near-wall region and follows the local streaming rolls adjacent to the boundary. At late times, it becomes trapped near a vortex centre located close to the wall, as shown in  Fig.~\ref{fig:fig06}(h). These results underscore the importance of considering both early- and late-time dynamics in acoustofluidic systems, as the trapping behaviour can evolve substantially over time.

\begin{figure}
\centering
\includegraphics[clip,width=1\textwidth]{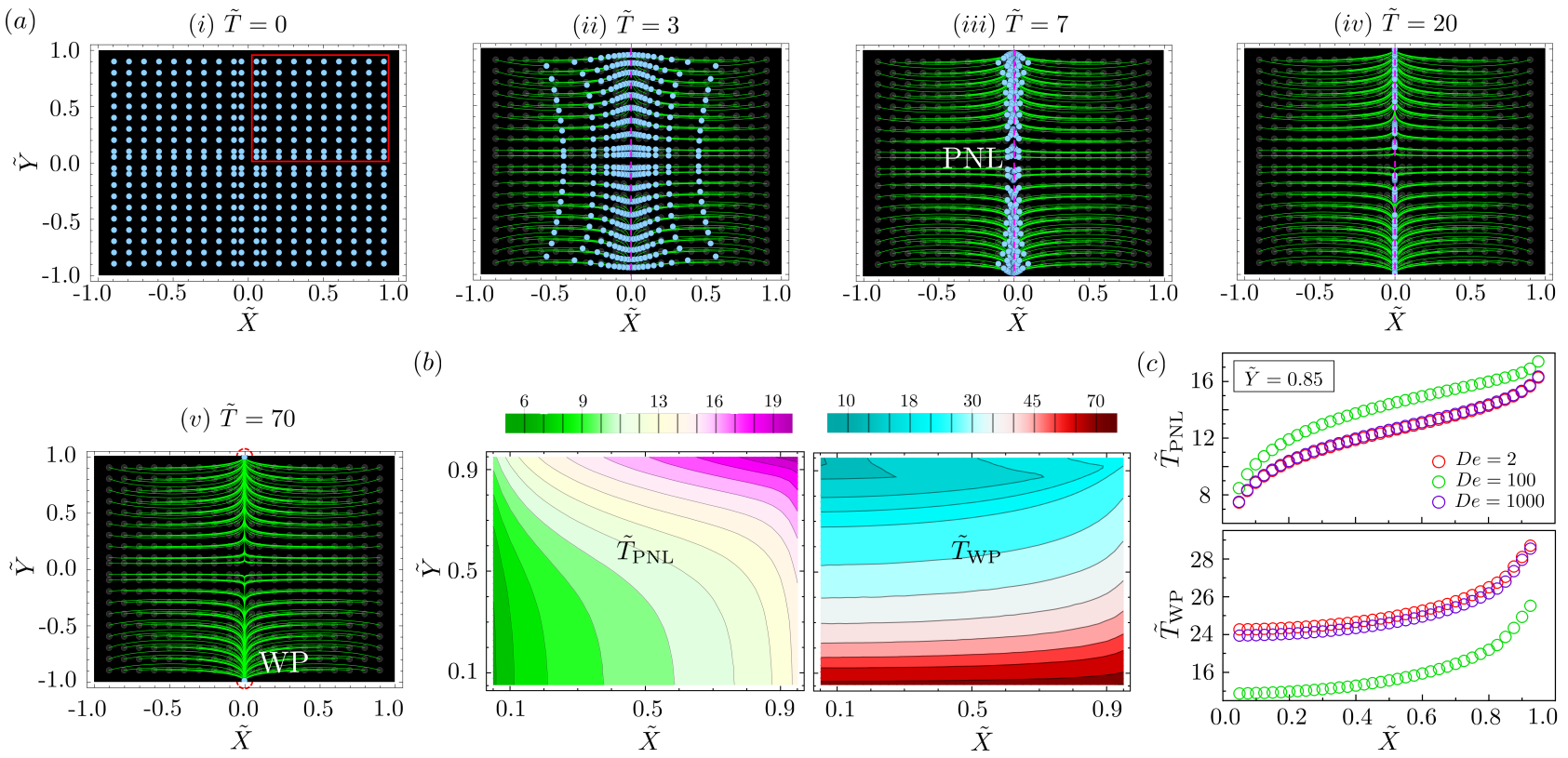}
\caption
{
\justifying{(a) Early- and late-time analysis of particle migration in the pressure-nodal-line to wall-point (PNL--WP) regime at $De=2$ and $Dv=20$. A single particle is initialized at different locations across the channel cross-section. 
(b) Spatial variation of the pressure-nodal trapping time, $\tilde{T}{\mathrm{PNL}}$, and wall-point trapping time, $\tilde{T}{\mathrm{WP}}$, across the channel cross-section at $De=2$ and $Dv=20$. 
(c) Variation of $\tilde{T}_{\mathrm{PNL}}$ and $\tilde{T}_{\mathrm{WP}}$ along $\tilde{X}$ at $\tilde{Y}=0.85$ for $De = 2,\ 100,$ and $1000$ at $Dv=20$.
 }
}
\label{fig:fig07}
\end{figure}

We next examine the early- and late-time dynamics in the PNL–WP regime. Figure~\ref{fig:fig07}(a) shows time-resolved particle positions during migration for $De=2$ and $Dv=20$. At early times, the particle is driven toward the pressure nodal line (PNL). At later times, however, the streaming-induced drag redirects the particle toward the wall point (WP), where it ultimately gets trapped. This behaviour is consistent with the experimental observations shown in Fig.~\ref{fig:Fig_7_3}-\ref{fig:Fig_7_5}. To quantify these dynamics, we define $\tilde{T}_\textrm{PNL}$ as the time required for a particle to reach the PNL and  $\tilde{T}_\textrm{WP}$ as the time required to reach the WP, from its initial position. Particles initially located closer to the PNL reach it more rapidly, while those released near the channel corners require longer times (see Fig.~\ref{fig:fig07}(b)). Furthermore, particles initially located near the ultrasound symmetry line (USL) must traverse a greater distance to reach the WP compared with those near the ceiling or floor. Consequently, particles closer to the USL exhibit larger values of $\tilde{T}_\textrm{WP}$ than those near the channel boundaries (see Fig.~\ref{fig:fig07}(b)). It is also observed that $\tilde{T}_\textrm{WP} >> \tilde{T}_\textrm{PNL}$, indicating that trapping at the PNL occurs at early times, whereas WP trapping emerges at later stages of migration. As noted earlier, the PNL–WP regime reappears at higher Deborah numbers (see Fig.~\ref{fig:fig2}(b)). To examine this behaviour, we compare the variation of $\tilde{T}_\textrm{PNL}$ and $\tilde{T}_\textrm{WP}$ along $\tilde{X}$ at $\tilde{Y}=0.85$ for $De=2$, $100$, and $1000$, as shown in Fig.~\ref{fig:fig07}(c). At $De=100$, $\tilde{T}_\textrm{PNL}$ increases while $\tilde{T}_\textrm{WP}$ decreases relative to the case $De=2$. However, with a further increase to $De=1000$, this trend reverses: $\tilde{T}_\textrm{PNL}$ decreases and $\tilde{T}_\textrm{WP}$ increases. This non-monotonic behaviour arises from the variation of the streaming-induced drag force with $De$, $\hat{F}_{\textrm{str},\,De=100} > \hat{F}_{\textrm{str},\,De=1000} > \hat{F}_{\textrm{str},\,De=2}$ (refer inset in \autoref{appA} Fig. \ref{fig:fig019}). At $\tilde{Y}=0.85$, the streaming-induced drag force $\tilde{F}_\textrm{str}$ opposes the acoustic radiation force $\tilde{F}_\textrm{rad}$ during migration toward the PNL (see Fig.~\ref{fig:fig04}(a)). Consequently, an increase in streaming drag slows the migration toward the PNL, leading to larger $\tilde{T}_\textrm{PNL}$. In contrast, the same drag force accelerates migration toward the WP, resulting in smaller $\tilde{T}_\textrm{WP}$.
\smallskip

\begin{figure}
\centering
\includegraphics[clip,width=1\textwidth]{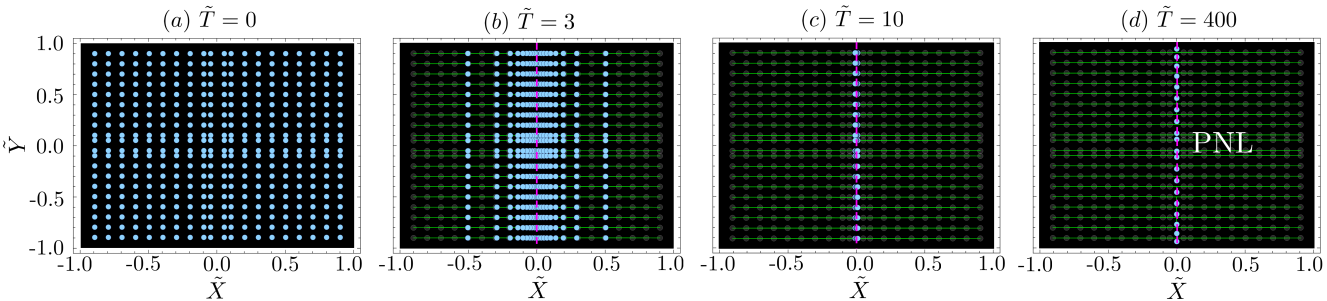}
\caption
{
\justifying{Early- and late-time analysis of particle migration in the pressure-nodal-line (PNL) regime at $De=2.05$ and $Dv=20$. A single particle is initialized at different locations across the channel cross-section. 
 }
}
\label{fig:fig08}
\end{figure}

\noindent \textit{Pressure nodal line trapping regime (PNL)}: We examine the trapping dynamics in the PNL regime at $De=2.05$ and $Dv=20$, as shown in Fig.~\ref{fig:fig08}. Here, particles migrate directly toward the pressure nodal line (PNL) and remain stably trapped there at both early and late times, without any subsequent migration toward the wall points. The particle trajectories are rectilinear and aligned with the nodal line, and all particles remain confined to the PNL at $\tilde{T}=400$.
\smallskip

\begin{figure}
\centering
\includegraphics[clip,width=1\textwidth]{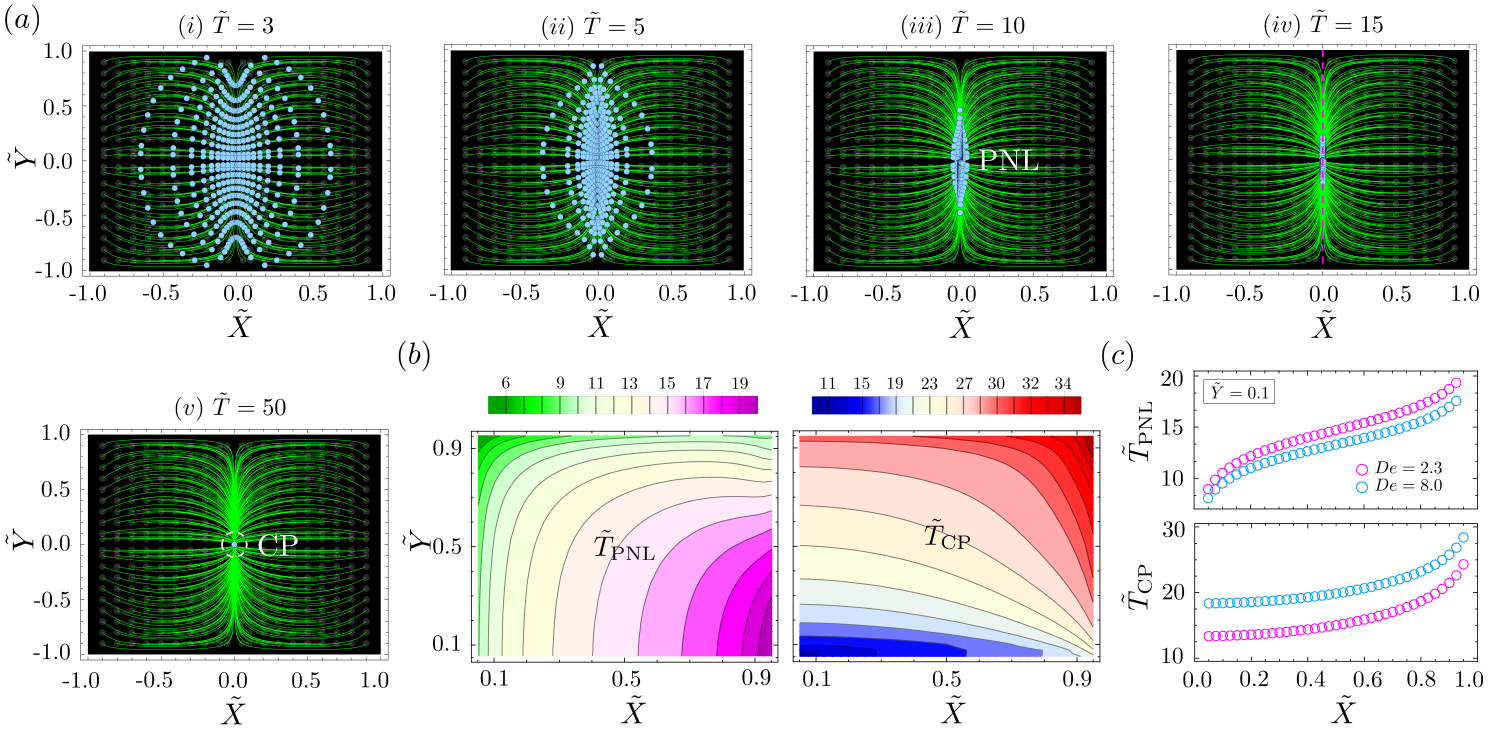}
\caption
{
\justifying{(a) Early- and late-time analysis of particle migration in the pressure-nodal-line to center-point (PNL--CP) regime at $De=2.3$ and $Dv=20$. A single particle is initialized at different locations across the channel cross-section. 
(b) Spatial variation of the pressure-nodal trapping time, $\tilde{T}{\mathrm{PNL}}$, and center-point trapping time, $\tilde{T}{\mathrm{CP}}$, across the channel cross-section at $De=2.3$ and $Dv=20$.
(c) Variation of $\tilde{T}_{\mathrm{PNL}}$ and $\tilde{T}_{\mathrm{CP}}$ along $\tilde{X}$ at $\tilde{Y}=0.1$ for $De = 2.3$ and $8$ at $Dv=20$.
 }
}
\label{fig:fig09}
\end{figure}

\noindent \textit{Pressure nodal line to centre point trapping regime (PNL-CP)}:  Beyond the PNL regime, the ratio $\hat{F}_\textrm{rad}/\hat{F}_\textrm{str}$  becomes negative (see Fig.~\ref{fig:fig04}), indicating a reversal in the direction of the streaming-induced drag force. The early- and late-time dynamics of the PNL–CP regime at $De=2.3$ and $Dv=20$ are shown in Fig.~\ref{fig:fig09}. Here, the competition between the acoustic radiation force $\tilde{F}_\textrm{rad}$ and the reversed streaming-induced drag $\tilde{F}_\textrm{str}$ produces trajectories that curve toward the channel centre, located near the pressure nodal line (PNL). Consequently, the particle distribution exhibits a convex trapping pattern at early times. As time progresses, the reversed streaming drag, directed toward the channel centre, progressively transports particles inward. At late times ($\tilde{T}=50$), this inward motion leads to stable trapping at the centre point (CP), as shown in Fig.~\ref{fig:fig09}(a)-v. To quantify the migration dynamics, we examine the variation of trapping times required for particles to reach PNL ($\tilde{T}_\textrm{PNL}$) and CP ($\tilde{T}_\textrm{CP}$) across one quarter of the channel cross-section (see Fig.~\ref{fig:fig09}(b)). Particles initially located closer to the PNL reach the nodal line more rapidly, whereas those farther away require longer migration times. A similar trend is observed for the trapping time to the CP, denoted by $\tilde{T}_\textrm{CP}$: particles located near the channel centre reach the CP more quickly, while those farther away take longer due to the increased migration distance. The PNL–CP regime reappears at higher elasticity, for example at $De=8$ (refer \autoref{appA} Fig. \ref{fig:fig019}). Under these conditions, the streaming-induced drag is reversed, and in the vicinity of the ultrasound symmetry line (USL) it acts opposite to the acoustic radiation force. To illustrate this behaviour, we compare the variations of $\tilde{T}_\textrm{PNL}$ and $\tilde{T}_\textrm{CP}$ for $De=2.3$ and $De=8$ at $\tilde{Y}=0.1$, as shown in Fig.~\ref{fig:fig09}(c). Owing to the larger magnitude of the streaming-induced drag force, \(\hat{F}_{\mathrm{str}}\), at \(De=2.3\) compared to \(De=8\), particles require a longer time to migrate toward the PNL, resulting in higher values of \(\tilde{T}_{\mathrm{PNL}}\) (see \autoref{appA} inset of Fig \ref{fig:fig019}). In contrast, the same enhanced drag accelerates migration toward the CP, leading to reduced values of $\tilde{T}_\textrm{CP}$ compared with the case $De=8$.
\smallskip


\begin{figure}
\centering
\includegraphics[clip,width=1\textwidth]{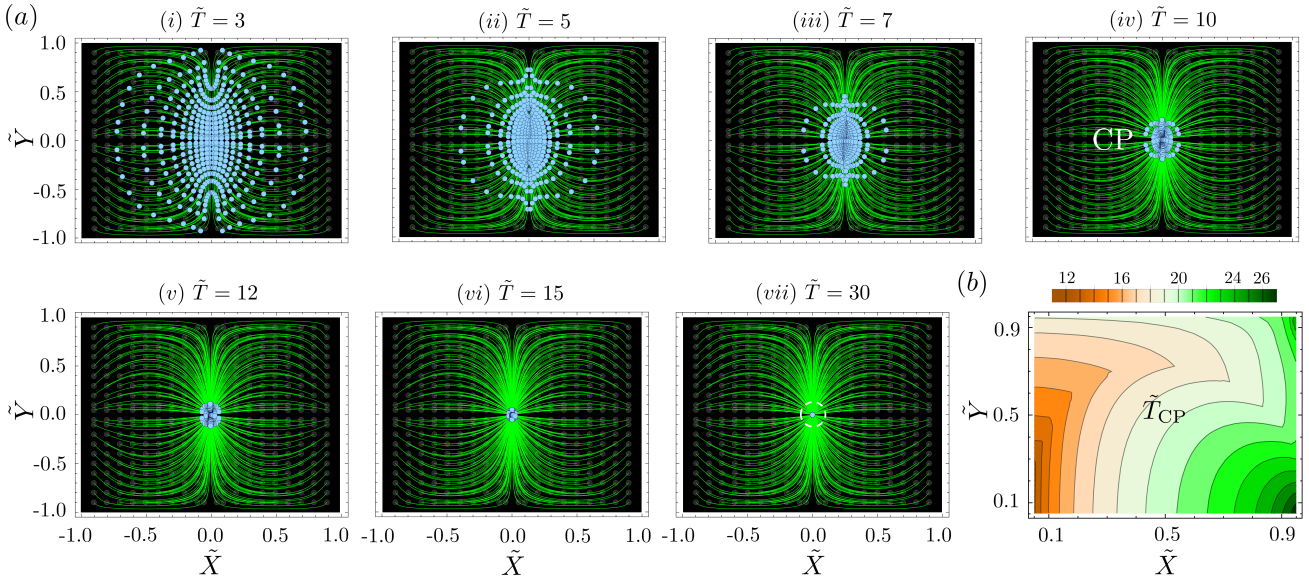}
\caption
{
\justifying{(a) Early- and late-time analysis of particle migration in the center-point (CP) regime at $De=2.5$ and $Dv=20$. A single particle is initialized at different locations across the channel cross-section. 
(b) Spatial variation of the center point trapping time, $\tilde{T}_{\mathrm{CP}}$, across the channel cross-section at $De=2.5$ and $Dv=20$.
 }
}
\label{fig:fig010}
\end{figure}

\begin{figure}
\centering
\includegraphics[clip,width=1\textwidth]{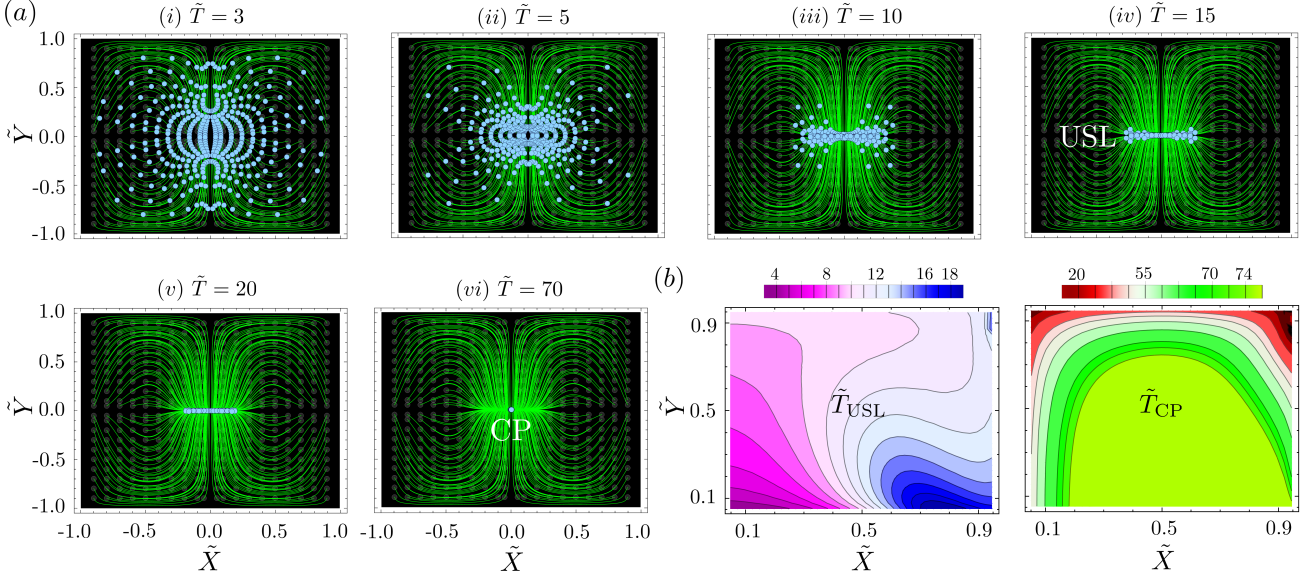}
\caption
{
\justifying{(a) Early- and late-time analysis of particle migration in the ultrasound symmetry line--center point (USL-CP) regime. A single particle is initialized at different locations across the channel cross-section at $De=4$ and $Dv=20$. (b) Spatial variation of the ultrasound symmetry-line trapping time, $\tilde{T}{\mathrm{USL}}$, and center-point trapping time, $\tilde{T}{\mathrm{CP}}$, across the channel cross-section at $De=4$ and $Dv=20$. 
 }
}
\label{fig:fig011}
\end{figure}

\noindent \textit{Centre-point trapping regime (CP)}: We next examine the early- and late-time dynamics of the centre point (CP) regime at $De=2.5$ and $Dv=20$ (see Fig.~\ref{fig:fig010}). Here, particles follow curved trajectories toward the channel centre during the early stages of migration and ultimately become trapped there. The reversed streaming-induced drag dominates over the acoustic radiation force, driving particles directly toward the CP. As a result, the trapping pattern assumes a circular distribution centred about the channel axis. The variation of the trapping time at the channel centre, denoted by $\tilde{T}_\textrm{CP}$, is shown in Fig.~\ref{fig:fig010}(b). Particles initially located closer to the pressure nodal line (PNL) reach the CP more rapidly than those farther away. This behaviour arises because the streaming-induced drag directed toward the CP is stronger near the PNL, thereby accelerating particle migration toward the trapping location.
\smallskip

\noindent \textit{Ultrasound symmetry line to centre point trapping regime (USL-CP)}: We next examine the early- and late-time dynamics of the USL–CP regime at $De=4$ and $Dv=20$ (see Fig.~\ref{fig:fig011}). Here, particles initially migrate toward the ultrasound symmetry line (USL) and undergo transient trapping during the early stage of motion. As time progresses, they depart from the USL and migrate toward the channel centre, where stable trapping is achieved at late times. The variations of the migration times toward the USL and the CP, denoted by $\tilde{T}_\textrm{USL}$ and $\tilde{T}_\textrm{CP}$, respectively, are shown in Fig.~\ref{fig:fig011}(b). In this regime, the force minimum is localised along the USL in the vicinity of the CP, rather than being distributed along its entire extent. Consequently, early-time trapping along the USL is spatially non-uniform. Particles initially located closer to the pressure nodal line (PNL) reach the USL more rapidly, resulting in smaller values of $\tilde{T}_\textrm{USL}$, whereas those farther away require longer migration times. A notable feature of this regime is the spatial redistribution of particles along the USL during the early stage of migration. Particles initially located farther from the USL approach it at positions closer to the CP, while those initially nearer to the USL tend to settle farther from the CP along the USL. This redistribution influences the subsequent migration toward the CP; particles that reach the USL closer to the CP traverse a shorter distance and therefore reach the centre more rapidly. Consequently, such particles exhibit smaller values of $\tilde{T}_\textrm{CP}$, as shown in Fig.~\ref{fig:fig011}(b).

\begin{figure}
\centering
\includegraphics[clip,width=1\textwidth]{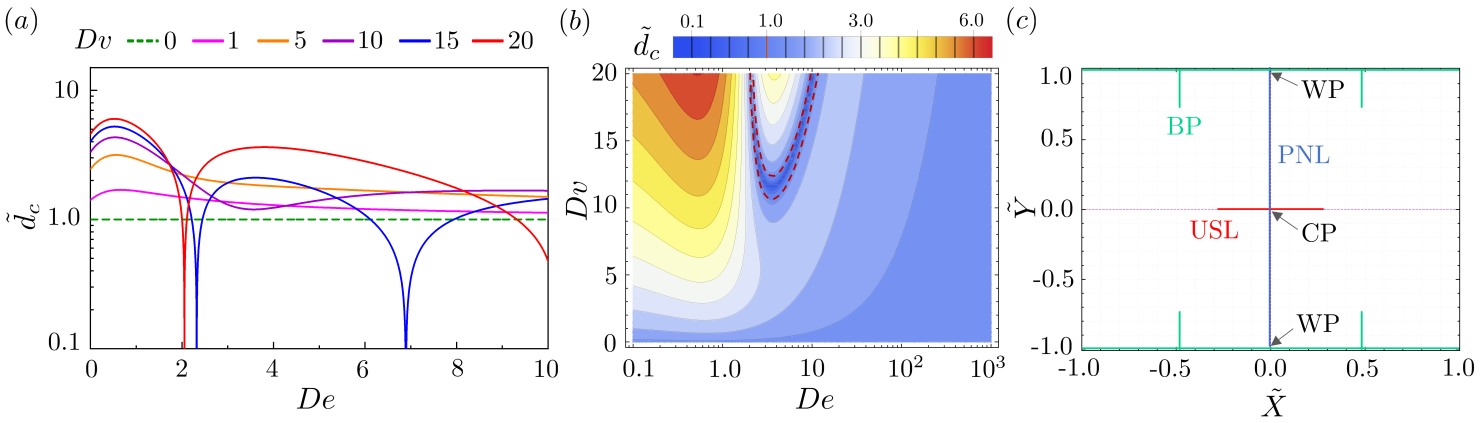}
\caption
{
\justifying{(a) Variation of the normalized critical particle diameter $\tilde{d}_c$ with Deborah number ($De$) at different viscous diffusion number ($Dv$). (b) Contour representation normalized critical particle diameter $\tilde{d}_c$ with $De$ and $Dv$. (c) Variation of  particle trapping locations across the cross section of channel. Here, BP, WP, CP, PNL, and USL denote bulk point, wall point, center point, pressure nodal line, and ultrasound symmetry line, respectively.
 }
}
\label{fig:fig012}
\end{figure}

\subsection{Critical particle size for acoustic migration and trapping in viscoelastic fluids }\label{sec:4.4}

Understanding the migration dynamics of micron and submicron-sized particles under ultrasound in viscoelastic fluids is important for a wide range of microfluidic applications. In such systems, the transition from streaming-dominated to radiation-dominated acoustophoresis is governed by a critical particle size. When $a \gg a_c$, particle motion is dominated by the acoustic radiation force. In contrast, when the particle size approaches or falls below the critical size ($a \lesssim a_c$), acoustic streaming becomes significant, and the migration dynamics are strongly influenced by the viscoelastic properties of the fluid. We identify a corresponding critical particle size for viscoelastic fluids that incorporates both elastic and viscous effects, as defined in Eq. (\ref{eq:0_46}). To facilitate comparison with the Newtonian limit, we introduce a dimensionless critical size, $\tilde{a}_c = a_c^{ve}/a_c^{bf}$, where $a_c^{bf}$ denotes the critical size of the base (Newtonian) fluid. The variation of $\tilde{d}_c=2\tilde{a}_c$ with the Deborah number ($De$) and viscous diffusion number ($Dv$) is shown in Fig.~\ref{fig:fig012}. For high $Dv$ ($=15$ and $20$), $\tilde{d}_c$ exhibits a non-monotonic dependence on $De$, reflecting the alternating strengthening and weakening of streaming effects. Remarkably, at specific values of $De$, $\tilde{d}_c$ becomes significantly smaller than unity, suggesting that viscoelasticity enables effective manipulation of submicron particles within microchannels. This behaviour is consistent with the competition between acoustic radiation and streaming-induced drag forces discussed in Fig.~\ref{fig:fig04}. For lower $Dv$ ($=10,\,5,\,1,$ and $0$), $\tilde{d}_c$ initially increases with $De$ and subsequently decreases, again governed by the evolving balance between radiation and streaming forces. A contour map of $\tilde{d}_c$ is presented in Fig.~\ref{fig:fig012}(b), delineating the parameter space for controlled manipulation and trapping of particles across a wide range of sizes, depending upon $De$ and $Dv$.

The different possible particle trapping regimes are summarised in Fig.~\ref{fig:fig012}(c). Depending on the viscoelastic properties and acoustic parameters, particles can be directed and trapped along the pressure nodal line (PNL), the ultrasound symmetry line (USL), at the channel centre (CP), at the bulk (BP), or near the channel walls (WP). Overall, this study establishes a unified framework linking force competition, viscoelasticity, and particle size to controllable migration and trapping under ultrasound. The ability to tune trapping locations through fluid elasticity and acoustic actuation provides a versatile mechanism for particle focusing, separation, and sorting in complex fluids. These findings are particularly relevant to lab-on-a-chip platforms involving biological samples, where fluid rheology plays a central role, and offer a pathway toward advanced acoustofluidic manipulation of cells, microorganisms, and submicron particles in biomedical and diagnostic applications.

\section{Conclusions}\label{sec:6}

We investigated the combined effects of acoustic radiation forces (ARF) and acoustic streaming (AS) on particle migration in viscoelastic fluids. While ARF drives particles toward pressure nodes or antinodes, streaming-induced flows can modify or oppose this motion. The resulting competition governs particle trajectories and trapping behaviour within the microchannel. Our results demonstrate that fluid viscoelasticity provides a powerful and independent control parameter for acoustophoretic manipulation. By tuning the viscoelastic properties, the particle trajectories can be systematically modulated, enabling controlled trapping at the pressure nodal line, channel walls, bulk regions, channel centre or the ultrasound symmetry line. This tunability arises from the dependence of the force balance on two key dimensionless parameters: the viscous diffusion number, $Dv$, and the Deborah number, $De$, which together regulate the relative strength and direction of ARF and AS. An early- and late-time analysis reveals that particle migration is inherently dynamic, with trajectories and trapping locations evolving over time. Depending on $De$ and $Dv$, multiple migration and trapping regimes emerge. These results highlight the critical role of viscoelasticity in shaping both transient and steady-state particle behaviour. The theoretical predictions are supported by experiments, which capture the transitions between radiation-dominated and streaming-dominated regimes and confirm the predicted trapping locations. We further determine the critical particle size in viscoelastic fluids and show that the critical particle diameter can be significantly smaller than that of the corresponding Newtonian fluid. This reduction enables effective manipulation of submicron particles, which are typically difficult to control using conventional acoustofluidic approaches. In addition, the viscoelastic modulation of force balance allows for the differentiation of particles with the same sign of acoustic contrast factor, a capability not readily achievable in Newtonian systems. Overall, the present study establishes a unified framework linking force competition, viscoelasticity, and particle size to controllable migration and trapping under ultrasound. These findings open new possibilities for precise, geometry-independent particle manipulation in complex fluids, with potential applications in lab-on-a-chip systems and biomedical technologies involving handling of cells, microorganisms, and submicron entities.

\backsection[Acknowledgements]{ A.K.S. acknowledges funding support from Anusandhan
National Research Foundation (ANRF), Government of India, through Grant No. ANRF/ARG/2025/000464/ENS, and Ministry of Education, Government of India, under IoE Phase II project. The authors also acknowledge support of CNNP, IIT Madras for the device fabrication.}


\backsection[Declaration of interests]{The authors report no conflict of interest.}

\appendix

\section{Streaming-induced drag force variation with Deborah number}\label{appA}

\begin{figure}
\centering
\includegraphics[clip,width=0.6\textwidth]{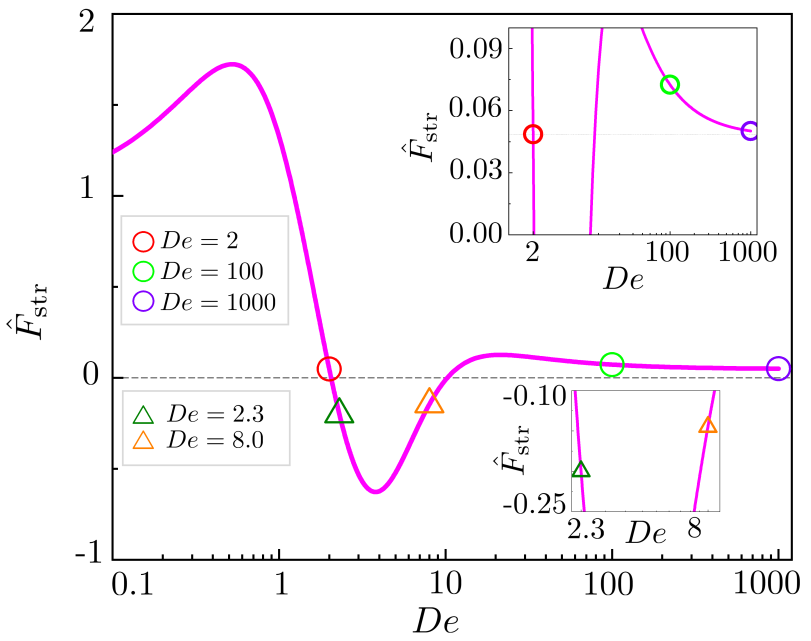}
\caption
{
\justifying{Variation of the dimensionless streaming-induced drag force, \( \hat{F}_{\mathrm{str}} \), with the Deborah number, \( De \), for \( Dv=20 \). The top inset compares the representative force magnitudes at \( De=2 \), \(100\), and \(1000\), showing that the force at \( De=100 \) is higher than those at \( De=2 \) and \( De=1000 \). The bottom inset compares the representative force magnitudes at \( De=2.3 \) and \(8\), showing that the magnitude of force at \( De=2.3 \) is higher than those at  \( De=8 \).
 }
}
\label{fig:fig019}
\end{figure}

The streaming-induced drag force exhibits a non-monotonic variation with the Deborah number, as shown in the Fig. \ref{fig:fig019}. Initially, \( \hat{F}_{\mathrm{str}} \) increases with \( De \), reaches a maximum, and then decreases sharply with further increase in \( De \). At intermediate \( De \), the force becomes negative due to the reversal of the streaming flow direction. With a further increase in \( De \), \( \hat{F}_{\mathrm{str}} \) becomes positive again and shows a slight increase before gradually decreasing at very high \( De \). For the representative cases shown in the inset, \( De=2 \) corresponds to the transition region where the streaming-induced drag force is close to zero. At \( De=100 \), the force becomes positive again and attains a relatively higher magnitude. Further increasing the Deborah number to \( De=1000 \) leads to a slight reduction in the force magnitude, indicating a weak decay of the streaming-induced drag force in the high-\( De \) regime. Here from Fig. \ref{fig:fig019} top inset, $\hat{F}_{\textrm{str},De=100}>\hat{F}_{\textrm{str},De=1000}>\hat{F}_{\textrm{str},De=2}$. The points corresponding to \(De=2.3\) and \(De=8\) denote the reversal of the streaming direction. The bottom inset compares the corresponding force amplitudes, showing that the force magnitude at \(De=2.3\) is greater than that at \(De=8\), i.e., $|F_{\mathrm{str}}|_{De=2.3} > |F_{\mathrm{str}}|_{De=8}$.

\counterwithin*{equation}{section}
\renewcommand\theequation{\thesection\arabic{equation}}

\bibliographystyle{jfm}
\bibliography{jfm}

\end{document}